\documentclass[
  aps,
  prc,
  twocolumn,
  superscriptaddress,
  nofootinbib,
  longbibliography,
  floatfix
]{revtex4-2}

\usepackage[T1]{fontenc}
\usepackage{amsmath,amssymb,bm}
\usepackage{graphicx}
\usepackage{tikz}
\usepackage{booktabs}
\usepackage{xcolor}
\usepackage{hyperref}
\usepackage{slashed}
\usepackage{comment}

\hypersetup{colorlinks=true,citecolor=blue,linkcolor=blue,urlcolor=blue}

\newcommand{\unit}{\mathbf{1}}

\begin{document}

\title{Antinucleon--nucleon interaction at next-to-next-to-leading order in covariant chiral effective field theory}

\author{Yang Xiao}
\affiliation{School of Space and Environment, Beihang University,
Beijing 102206, China}
\affiliation{School of Physics, Beihang University, Beijing 102206, China}
\affiliation{Center for Exotic Nuclear Studies, Institute for Basic Science, Daejeon 34126, Republic of Korea}

\author{Ye-Bin Wen}
\affiliation{School of Physics, Beihang University, Beijing 102206, China}

\author{Jun-Xu Lu}
\email[Corresponding author: ]{ljxwohool@buaa.edu.cn}
\affiliation{School of Physics, Beihang University, Beijing 102206, China}

\author{Youngman Kim}
\email[Corresponding author: ]{ykim@ibs.re.kr}
\affiliation{Center for Exotic Nuclear Studies, Institute for Basic Science, Daejeon 34126, Republic of Korea}

\author{Li-Sheng Geng}
\email[Corresponding author: ]{lisheng.geng@buaa.edu.cn}
\affiliation{Sino-French Carbon Neutrality Research Center, \'{E}cole Centrale de P\'{e}kin/School
of General Engineering, Beihang University, Beijing 100191, China}
\affiliation{School of Physics, Beihang University, Beijing 102206, China}
\affiliation{Peng Huanwu Collaborative Center for Research and Education,
Beihang University, Beijing 100191, China}
\affiliation{Southern Center for Nuclear-Science Theory (SCNT), Institute
of Modern Physics, Chinese Academy of Sciences, Huizhou 516000, China}

\date{\today}

\begin{abstract}
We construct antinucleon--nucleon ($\bar NN$) interactions at next-to-leading order (NLO) and next-to-next-to-leading order (NNLO) in covariant chiral effective field theory.  The short-range interaction is represented by complex covariant contact terms, while the long-range part is obtained from the corresponding nucleon--nucleon interaction through $G$-parity transformation. The low-energy constants are determined from the Nijmegen partial-wave analysis, and phase shifts and mixing angles are calculated for all partial waves with $J\leq2$.  Compared with NLO, NNLO improves the overall description of the phase shifts and mixing angles, particularly in several $J=2$ channels, although the improvement is not uniform across all partial waves or at higher energies.  We estimate Bayesian truncation uncertainties and compare the results with heavy-baryon chiral interactions.  We also revisit the pole candidates identified in our leading-order study and calculate the corresponding threshold parameters.  The isoscalar ${}^{13}P_0$ pole remains a few MeV below the $\bar NN$ threshold with a width of about 40 MeV at both NLO and NNLO, whereas the $S$-wave candidates show stronger sensitivity to chiral orders.  The NNLO covariant $\bar NN$ interaction therefore provides a useful basis for studies of antinucleon--nucleon scattering and near-threshold phenomena.
\end{abstract}

\maketitle

\section{Introduction}
\label{sec:introduction}

Structures near the proton–antiproton threshold have long motivated investigations of the antinucleon–nucleon interaction. They appear in charmonium decays, baryonic \(B\) decays, and  \(e^+e^-\to p\bar p\) reactions close to threshold~\cite{BES:2003aic,BES:2005ega,BESIII:2011aa,BESIII:2016fbr,BESIII:2021xoh,Belle:2002bro,Belle:2002fay,BaBar:2005sdl,BaBar:2005pon}. Recent measurements include the anomalous \(X(1840)\) line shape and the \(X(1880)\) structure reported by the BESIII Collaboration~\cite{BESIII:2023vvr}, as well as threshold enhancements in baryonic heavy-hadron decays~\cite{LHCb:2025ppbar}. The dynamical origin of these structures remains unsettled.  Analyses incorporating $\bar NN$ final-state interactions have shown that rescattering can generate pronounced threshold enhancements and modify the corresponding line shapes~\cite{ZouChiang:2004OPEFSI,Kerbikov:2004Threshold,Sibirtsev:2004id,Haidenbauer:2006Timelike,Haidenbauer:2014kja,Dmitriev:2016RealVirtual,MilsteinSalnikov:2017Jpsi,Yang:2024x1880}. Baryonium and $\bar NN$ bound-state interpretations have also been proposed~\cite{DattaODonnell:2003Baryonium,Dedonder:2009bk,Yan:2004xs}, and near-threshold poles can themselves arise dynamically from $\bar NN$ rescattering~\cite{Niu:2024pole}. An isoscalar ${}^{3}P_0$ resonance was also found in a constituent quark model~\cite{EntemFernandez:2006QuarkModel}. Discriminating among these possibilities requires a controlled description of low-energy \(\bar NN\) amplitudes.

A characteristic feature of low-energy $\bar NN$ scattering is the strong coupling to annihilation channels, which at low energies are dominated by multi-meson final states. After these channels are integrated out, their effects at low momenta can be parametrized by the complex short-range components of the $\bar NN$ interaction~\cite{Kang:2013uia,Dai:2017ont,Xiao:2024nbarN}. Such absorptive effects have traditionally been incorporated through phenomenological optical potentials~\cite{Dover:1980pd,Dover:1981pp,Cote:1982gr,Hippchen:1991rr,Mull:1994gz,ElBennich:2009Paris}. Chiral effective field theory (ChEFT) provides a systematic framework based on the low-energy symmetries of QCD~\cite{Weinberg:1990rz,Weinberg:1991um,Weinberg:1992yk}. The pion-exchange contributions are related to their \(NN\) counterparts by \(G\) parity, whereas unresolved short-distance interactions and annihilation effects are encoded in contact terms. Early EFT studies included explicit pion exchanges and represented annihilation through complex short-range couplings~\cite{Chen:2010NbarNEFT,ChenMa:2011NLO}. Within a Weinberg-type power counting, heavy-baryon (HB) ChEFT $\bar NN$ interactions were constructed up to NNLO in Ref.~\cite{Kang:2013uia} and extended to N$^3$LO in Ref.~\cite{Dai:2017ont}. These interactions provide a quantitative description of the energy-dependent Nijmegen partial-wave analysis (PWA)~\cite{Zhou:2012ui}. More recently, an SU(3) ChEFT analysis of $\bar pp\to\bar\Lambda\Lambda$ employed chiral meson-exchange interactions supplemented by a phenomenological optical potential to account for annihilation effects~\cite{Yang:2026barB}. The renormalization of iterated one-pion exchange (OPE) has been investigated in low and higher partial waves~\cite{Zhou:2022loi,Zhou:2023HigherWaves}. These interactions have been used in analyses of near-threshold production and timelike electromagnetic form factors~\cite{Haidenbauer:2014kja,Kang:2015yka,Haidenbauer:2015yka,Dai:2018tlc,Yang:2022qoy,Yang:2024iuc,Yang:2024x1880,Ji:2026mfe}.

The HB expansion provides a well-established framework for nuclear forces. A covariant formulation retains the Dirac structure of the baryons and relativistic kinematics, leading to a different organization of relativistic contributions and contact-term momentum dependence at a fixed chiral order. Covariant baryon ChEFT~\cite{Gasser:1987rb,Fuchs:2003EOMS} has been applied extensively in the one-baryon and meson--baryon sectors~\cite{Geng:2008mf,Ren:2012aj,Alarcon:2011zs,Lu:2022hwm}.  The covariant $NN$ interaction in this framework has been developed systematically from LO up to NNLO~\cite{Ren:2016jna,Ren:2017yvw,Xiao:2018jot,Xiao:2020ozd,Wang:2021kos,Lu:2021gsb,Xiao:2025ufy}. For recent reviews of relativistic chiral nuclear forces, see Refs.~\cite{Lu:2024review,Lu:2025review}. Recently, it was employed in relativistic Brueckner--Hartree--Fock calculations of nuclear matter~\cite{Zou:2023quo,Zou:2025dao} and finite nuclei~\cite{Shen:2025iue}.
It has also been successfully used in studies of in-medium hyperon--nucleon forces~\cite{Zheng:2025lambdaN} and finite hypernuclei~\cite{Zeng:2025fqb}.

We have applied the covariant framework to the $\bar NN$ system at LO~\cite{Xiao:2024nbarN}. The LO interaction of Ref.~\cite{Xiao:2024nbarN}, however, employed a hybrid treatment of the short-range dynamics. The elastic interaction was derived within covariant ChEFT, whereas annihilation was represented by a separately factorized short-range interaction. This factorized form implements the absorptive constraints explicitly~\cite{Kang:2013uia}. The resulting interaction describes many phase parameters at a level comparable to the HB NLO interaction. Nevertheless, higher-order corrections are needed to improve the description of the PWA. In addition, several pole candidates identified at LO displayed appreciable cutoff sensitivity, especially in the ${}^{11}S_0$ channel. It is therefore important to investigate how the scattering amplitudes and candidate poles evolve beyond LO. At the same time, a common covariant operator basis provides an alternative parametrization of the short-range dispersive and absorptive dynamics.

Accordingly, in the present work we construct the covariant $\bar NN$ interaction at NLO and NNLO. The long-range interaction contains the corresponding one- and two-pion-exchange contributions, while the short-range interaction is described by a common covariant contact basis with complex low-energy constants (LECs), whose real and imaginary parts parametrize dispersive and absorptive short-range contributions, respectively. At NNLO, subleading two-pion exchange contributions are considered while the contact-operator basis remains unchanged. The complex contact couplings are fitted directly to the Nijmegen PWA for partial waves with $J\leq2$. The resulting elastic $S$ matrices are checked numerically for $S^\dagger S \leq 1$ over the energy range considered. We calculate the phase shifts, inelasticities, and mixing angles, estimate conditional Bayesian truncation uncertainties from the calculated order-by-order contributions, and compare the results with the HB interaction. Finally, we revisit the pole candidates of Ref.~\cite{Xiao:2024nbarN}, calculate their scattering lengths and volumes and effective ranges, and assess their order dependence at a fixed cutoff.

The paper is organized as follows. The interaction and scattering equations are described in Secs.~\ref{sec:interaction} and \ref{sec:scattering}, respectively. The fitting strategy and uncertainty estimates are presented in Sec.~\ref{sec:numerical}. Scattering observables are discussed in Sec.~\ref{sec:phase shifts}, followed by the pole analysis in Sec.~\ref{sec:poles}. We summarize our conclusions in Sec.~\ref{sec:summary}.

\section{Covariant $\bar NN$ interaction at NNLO}
\label{sec:interaction}

\begin{figure*}[t]
  \centering
  \includegraphics[width=\textwidth]{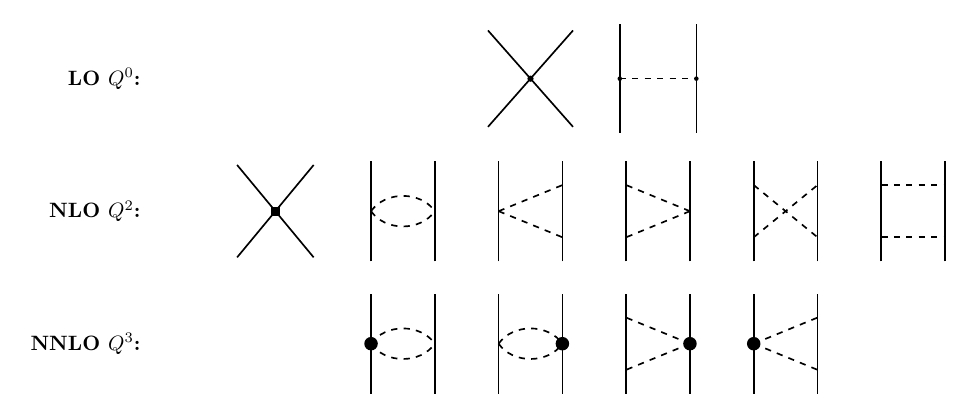}
  \caption{$\bar N N$ Feynman diagrams up to NNLO. Solid and dashed lines denote nucleons/antinucleons and pions, respectively. The filled contact vertices originate from ${\cal L}_{NN}^{(0,2)}$, while the small and enlarged filled circles denote vertices from ${\cal L}_{\pi N}^{(1)}$ and ${\cal L}_{\pi N}^{(2)}$, respectively.}
  \label{fig:diagrams}
\end{figure*}

In a covariant formulation, the external antinucleon is represented by a Dirac $v$ spinor, rather than the $u$ spinor used for a nucleon.  Nevertheless, the generalized Fierz identities allow four-baryon bilinears with different $u$- and $v$-spinor orderings to be rearranged into the same covariant operator basis~\cite{Nieves:2003in,Xiao:2024nbarN}.  This rearrangement establishes the equivalence of the Dirac operator bases, but does not by itself provide an independent chiral power counting for the eliminated annihilation channels.  The classification of two-body topologies is the same as in the $NN$ sector, while the differences reside in the external spinors and the $G$-parity transformation of the pion-exchange contributions.  Fig.~\ref{fig:diagrams} organizes the contributions retained in the present work by chiral order. Here $Q$ denotes a generic soft scale of ChEFT, such as an external three-momentum or the pion mass, measured relative to the chiral breakdown scale. At LO ($Q^0$), the potential contains a nonderivative contact interaction and the one-pion exchange (OPE).  At NLO ($Q^2$), the two-derivative contact interactions and the five leading two-pion-exchange (TPE) topologies enter.  At NNLO ($Q^3$), no new contact operator appears, while the TPE diagrams contain one insertion
from the subleading $\pi N$ Lagrangian ${\cal L}_{\pi N}^{(2)}$.

The elastic contact and pion-exchange contributions are derived from the same covariant chiral Lagrangians as in the $NN$ sector~\cite{Ren:2016jna,Xiao:2018jot,Wang:2021kos,Lu:2021gsb}.  The corresponding crossed $\bar NN$ elastic matrix elements are obtained by replacing one pair of external nucleon spinors by antinucleon spinors, schematically,
\begin{equation}
 [\bar u\,\Gamma_i u]\,[\bar u\,\Gamma_j u]
 \;\longrightarrow\;
 [\bar v\,\Gamma_i v]\,[\bar u\,\Gamma_j u],
 \label{eq:u-to-v}
\end{equation}
together with the appropriate $G$-parity transformation for the pion-exchange contributions.  Annihilation effects are incorporated below through the imaginary parts of effective complex contact coefficients.  The prescription for the elastic interaction is equivalent to that adopted in our LO $\bar NN$ study~\cite{Xiao:2024nbarN}.

The leading pion--nucleon interaction needed for OPE and the leading TPE contributions follows from
\begin{equation}
 {\cal L}_{\pi N}^{(1)}
 =
 \bar\Psi
 \left(
 i\slashed D-m_N
 +\frac{g_A}{2}\slashed u\gamma_5
 \right)\Psi ,
 \label{eq:LpiN1}
\end{equation}
where the standard chiral building blocks are defined as in Refs.~\cite{Wang:2021kos,Lu:2021gsb}. At NNLO, the subleading pion--nucleon Lagrangian relevant for two-pion-exchange contributions reads~\cite{Gasser:1987rb,Fettes:2000gb}
\begin{equation}
\begin{aligned}
 {\cal L}_{\pi N}^{(2)}
 ={}&
 c_1\langle\chi_+\rangle\,\bar\Psi\Psi -\frac{c_2}{4m_N^2}
 \langle u^\mu u^\nu\rangle
 \left(\bar\Psi D_\mu D_\nu\Psi+\mathrm{H.c.}\right)
 \\
 &+\frac{c_3}{2}
 \langle u_\mu u^\mu\rangle\,\bar\Psi\Psi -\frac{c_4}{4}
 \bar\Psi\gamma^\mu\gamma^\nu[u_\mu,u_\nu]\Psi .
\end{aligned}
\label{eq:LpiN2}
\end{equation}

The short-range interaction is generated by the covariant four-baryon contact Lagrangians,
\begin{equation}
 {\cal L}_{NN}^{(0)}
 =\sum_{i=1}^{4} C_i\,{\cal O}_i,
 \qquad
 {\cal L}_{NN}^{(2)}
 =\sum_{i=5}^{17} C_i\,{\cal O}_i ,
 \label{eq:Lct}
\end{equation}
where ${\cal O}_i$ denote the covariant four-baryon operators.  For example, the LO basis contains scalar structures of the form $(\bar\Psi\Psi)(\bar\Psi\Psi)$.  The complete operator basis and its partial-wave projection are given in Refs.~\cite{Xiao:2018jot,Lu:2021gsb}.  In the present $\bar NN$ interaction, the same operators are retained, with one nucleon spinor pair replaced by the corresponding antinucleon spinors as in Eq.~\eqref{eq:u-to-v}.

For fixed total isospin $I=0,1$, the NLO and NNLO potentials are written as
\begin{align}
 V_{\rm NLO}^{(I)}={}&V_{\rm CT}^{(0,I)}+V_{\rm CT}^{(2,I)}
 +V_{1\pi}^{(I)}+V_{2\pi}^{(2,I)}-V_{\rm IOPE}^{(I)},
 \label{eq:VNLO}\\
 V_{\rm NNLO}^{(I)}={}&V_{\rm NLO}^{(I)}+V_{2\pi}^{(3,I)}.
 \label{eq:VNNLO}
\end{align}
The superscripts $(0)$, $(2)$, and $(3)$ denote contributions of chiral orders $Q^0$, $Q^2$, and $Q^3$, respectively.

\subsection{Pion-exchange interaction}
\label{subsec:pions}

The pion-exchange part of the $\bar NN$ interaction follows directly from the corresponding $NN$ interaction by $G$ parity~\cite{Kang:2013uia,Dai:2017ont,Xiao:2024nbarN}.  An exchange of $n$ pions acquires the factor $(-1)^n$, so that
\begin{equation}
 V_{1\pi}^{\bar NN}=-V_{1\pi}^{NN},\qquad
 V_{2\pi}^{\bar NN}=+V_{2\pi}^{NN}.
 \label{eq:gparity}
\end{equation}
We therefore use the off-shell covariant $NN$ pion-exchange potentials of Ref.~\cite{Wang:2021kos} and apply Eq.~\eqref{eq:gparity}.  

For completeness, the Dirac spinors are defined as in our LO $\bar NN$ study~\cite{Xiao:2024nbarN},
\begin{align}
 u(\bm p,s)&=\mathcal N_p
 \begin{pmatrix}1\\[1pt] \bm\sigma\!\cdot\!\bm p/\epsilon_p\end{pmatrix}\chi_s,
 &\mathcal N_p&=\sqrt{\frac{\epsilon_p}{2m_N}},\nonumber\\[-1pt]
 \epsilon_p&=E_p+m_N,
 &E_p&=\sqrt{m_N^2+p^2},
 \label{eq:spinor-u}\\[-1pt]
 v(\bm p,s)&=\gamma^0 C\,u^*(\bm p,s),
 &C&=i\gamma^0\gamma^2.
 \label{eq:spinor-v}
\end{align}
Here $m_N$ is the nucleon mass, $s$ labels the spin projection, $\chi_s$ is a two-component Pauli spinor, $\bm\sigma$ denotes the Pauli matrices in spin space, $\gamma^\mu$ are the Dirac matrices, and $C$ is the
charge-conjugation matrix.  At LO, the OPE potential is sufficiently compact to display explicitly.  Adapting the OPE expression of Ref.~\cite{Wang:2021kos} to our isospin and metric conventions,
\begin{equation}
 V_{1\pi}^{NN}=
 \frac{g_A^2}{4f_\pi^2}\,
 \frac{[\bar u(p_3)\tau_1^a\gamma^\mu\gamma_5q_\mu u(p_1)]
       [\bar u(p_4)\tau_2^a\gamma^\nu\gamma_5q_\nu u(p_2)]}
      {q^2-m_\pi^2}.
 \label{eq:ope-wang}
\end{equation}
Here $g_A$ is the nucleon axial coupling, $f_\pi$ the pion decay constant, $m_\pi$ the pion mass, and $\tau_i^a$ the Pauli matrix in isospin space acting on baryon $i$; summation over the isospin index $a$ is understood.  In the c.m. frame, the external spinors are evaluated with $p_1=(E_p,\bm p)$ and $p_2=(E_p,-\bm p)$ for the incoming state and $p_3=(E_{p'},\bm p')$ and $p_4=(E_{p'},-\bm p')$ for the outgoing state. The exchanged pion carries $q=(0,\bm p-\bm p')$.  The $\bar NN$ OPE potential is obtained from Eq.~\eqref{eq:ope-wang} by replacing the antinucleon line by the corresponding $v$ spinors and applying the minus sign in Eq.~\eqref{eq:gparity}.

For TPE we use the covariant off-shell expressions of Ref.~\cite{Wang:2021kos}.  At fixed total c.m. energy $W$, the loop amplitudes are evaluated with the equal-energy prescription $p_i^0=W/2$ and $q^0=0$.  At NLO ($Q^2$), the TPE potential contains the planar box ($B$), crossed box ($X$), football ($F$), left triangle ($T_L$), and right triangle ($T_R$) diagrams,
\begin{equation}
 V_{2\pi}^{(2,I)}=
 \sum_{i=B,X,F,T_L,T_R}{\cal I}_i^{(I)}V_i^{(2)}.
 \label{eq:tpeQ2}
\end{equation}
Here $I=0,1$ is the total isospin of the $\bar NN$ system and ${\cal I}_i^{(I)}$ denotes the corresponding isospin factor for topology $i$.  The explicit off-shell potentials are given in Eqs.~(9)--(13) of Ref.~\cite{Wang:2021kos}, and their power-counting-breaking (PCB) pieces in Eqs.~(14)--(19).  We employ the extended-on-mass-shell (EOMS)-renormalized potential diagram by diagram,
\begin{equation}
 V_i^{(2),\mathrm{EOMS}}=V_i^{(2)}-V_{i,\mathrm{PCB}}^{(2)}.
 \label{eq:eomsQ2}
\end{equation}

At NNLO ($Q^3$), one vertex in a football or triangle diagram is replaced by a vertex from ${\cal L}_{\pi N}^{(2)}$.  We denote by $F_L$ ($F_R$) a football diagram with the subleading vertex on the left (right) baryon line, and by $T_L$ ($T_R$) the corresponding left- (right-) triangle diagram.  The NNLO TPE potential is therefore
\begin{equation}
 V_{2\pi}^{(3,I)}=
 \sum_{i=F_L,F_R,T_L,T_R}{\cal I}_i^{(I)}V_i^{(3)}.
 \label{eq:tpeQ3}
\end{equation}
The explicit $c_{1,2,3,4}$-dependent loop amplitudes are given in Eqs.~(20)--(29) of Ref.~\cite{Wang:2021kos}, and the corresponding PCB terms in Eqs.~(30)--(37).  The NNLO TPE potential used here is thus the EOMS-subtracted result
\begin{equation}
 V_{2\pi}^{(3),\mathrm{EOMS}}
 =V_{2\pi}^{(3)}-V_{2\pi,\mathrm{PCB}}^{(3)}.
 \label{eq:eomsQ3}
\end{equation}
The PCB pieces are polynomial in the external momenta and can be absorbed by the contact interaction~\cite{Wang:2021kos}.  No box or crossed-box topology appears at $Q^3$, because ${\cal L}_{\pi N}^{(2)}$ contains no one-pion $\pi NN$ vertex.  Eq.~\eqref{eq:gparity} then implies that the complete
TPE potential is unchanged when going from $NN$ to $\bar NN$.

Following the subtraction of the two-particle-reducible OPE iteration in Ref.~\cite{Wang:2021kos}, but adapting it to our partial-wave normalization, regulator, and Kadyshevsky propagator, we subtract
\begin{align}
 V_{{\rm IOPE},L'L}(p',p;W)={}&
 \sum_{L''}{\cal P}\!\int_0^\infty\frac{k^2dk}{(2\pi)^3}
 f_\Lambda^2(k)\nonumber\\
 &\times V_{1\pi,L'L''}(p',k;W)
 \frac{m_N^2}{2E_k^2}\nonumber\\
 &\times\frac{V_{1\pi,L''L}(k,p;W)}{W/2-E_k},
 \label{eq:iope}
\end{align}
where $L$, $L'$, and $L''$ denote orbital angular momenta in a fixed spin-angular-momentum channel, $p$ ($p'$) is the incoming (outgoing) c.m. momentum. $W$ is the total c.m. energy,  on shell, $W=2E_p$, $E_k=\sqrt{m_N^2+k^2}$, and ${\cal P}$ denotes the principal value.   The subtraction uses the same regulator and Kadyshevsky propagator as the scattering equation below.

\subsection{Complex contact interaction}
\label{subsec:contacts}

The short-range interaction is constructed from the same covariant contact operator basis as in Ref.~\cite{Lu:2021gsb}.  The standard partial-wave contact potentials up to NLO are not repeated here; their
explicit expressions are given in Eqs.~(14)--(22) of the Supplemental Material of Ref.~\cite{Lu:2021gsb}, together with the linear relations between the partial-wave coefficients and the 17 covariant operator-basis LECs.  There is no new contact operator at NNLO, and the same short-range basis is therefore refitted after the $Q^3$ pion-exchange contribution is included.

The essential extension to $\bar NN$ is that the short-range LECs are complex.  We write
\begin{equation}
 \mathcal C_a^{(I)}=
 \mathcal C_{a,R}^{(I)}+i\mathcal C_{a,I}^{(I)},
 \label{eq:complexLECs}
\end{equation}
where $a$ labels the covariant contact operator.  These coefficients parameterize the reduced short-range interaction after the open annihilation channels have been eliminated; they should not be interpreted as complex couplings of an underlying Hermitian Lagrangian.  The real and imaginary parts describe short-range elastic dynamics and annihilation effects, respectively.  Since several complex LECs contribute to a given partial wave, no sign is imposed on an individual imaginary coefficient; instead, the condition $S^\dagger S \leq 1$ of the resulting elastic $S$ matrix is checked numerically.

For later use, we define
\begin{equation}
 \Delta_p=E_p-m_N,\qquad N_p=E_p+m_N,
 \label{eq:contactdefs}
\end{equation}
with analogous definitions for $p'$.  Because the covariant contact operators correlate several partial waves, the nominal NLO basis is too restrictive for a simultaneous description of the $J=2$ amplitudes.  We therefore promote five nominally higher-order structures, following the same strategy used for the accurate covariant $NN$ interaction of Ref.~\cite{Lu:2021gsb},
\begin{align}
 \Delta V_{{}^1D_2}(p',p)&=
  \pi D_{{}^1D_2}\frac{p^2p'^2}{m_N^4},\label{eq:prom1D2}\\
 \Delta V_{{}^3D_2}(p',p)&=
  \pi D_{{}^3D_2}\frac{p^2p'^2}{m_N^4},
 \label{eq:prom3D2}\\
 \Delta V_{{}^3P_2}(p',p)&=
  \pi D_{{}^3P_2}\frac{pp'(\Delta_p+\Delta_{p'})}{m_N^3},
 \label{eq:promPP}\\
 \Delta V_{{}^3P_2\to{}^3F_2}(p',p)&=
  \pi D_{{}^3P_2-{}^3F_2}\frac{p\,p'^3}{m_N^3N_{p'}},\nonumber\\
 \Delta V_{{}^3F_2\to{}^3P_2}(p',p)&=
  \pi D_{{}^3P_2-{}^3F_2}\frac{p^3p'}{m_N^3N_p},
 \label{eq:promPF}\\
 \Delta V_{{}^3F_2}(p',p)&=
  \pi D_{{}^3F_2}\frac{pp'\Delta_p\Delta_{p'}}{m_N^4}.
 \label{eq:promFF}
\end{align}
These promoted structures belong to the covariant $\mathcal O(q^4)$ operator space constructed in Ref.~\cite{Xiao:2018jot}.

\section{Scattering equation and phase conventions}
\label{sec:scattering}

Partial waves are denoted by ${}^{2I+1,\,2S+1}L_J$, where $I$, $S$, $L$, and $J$ are the total isospin, spin, orbital angular momentum, and total angular momentum, respectively. The partial-wave amplitude is obtained by solving the Kadyshevsky equation in the $LSJ$ basis~\cite{Kadyshevsky:1968,Xiao:2024nbarN},
\begin{align}
 T_{L'L}^{SJ}(p',p;W)={}&V_{L'L}^{SJ}(p',p;W)\nonumber\\
 &+\sum_{L''}\int_0^\infty\frac{k^2dk}{(2\pi)^3}
 V_{L'L''}^{SJ}(p',k;W)\nonumber\\
 &\quad\times\frac{m_N^2}{2E_k^2}
 \frac{T_{L''L}^{SJ}(k,p;W)}{W/2-E_k+i0^+}.
 \label{eq:kadyshevsky}
\end{align}
The potential is regularized as
\begin{align}
 V_\Lambda(p',p)&=f_\Lambda(p')V(p',p)f_\Lambda(p),\nonumber\\
 f_\Lambda(p)&=\exp\!\left[-\left(\frac{p}{\Lambda}\right)^6\right].
 \label{eq:regulator}
\end{align}
We retain the same nonlocal Gaussian regulator used in our LO study. We have also checked at LO that replacing it with a semilocal regulator of the type advocated in Ref.~\cite{Reinert:2017usi} leads to only insignificant changes in the fitted phase shifts.  A systematic study of regulator dependence in the covariant formulation is beyond the scope of the present work.

With this normalization, the on-shell $S$ matrix is
\begin{equation}
 S_{L'L}^{SJ}(p)=\delta_{L'L}
 -i\frac{p m_N^2}{8\pi^2E_p}T_{L'L}^{SJ}(p,p;2E_p).
 \label{eq:StoT}
\end{equation}
For an uncoupled channel, we define the complex phase shift by
\begin{equation}
 S=e^{2i\delta},\qquad
 \delta=\delta_R+i\delta_I,
 \label{eq:uncoupledphase}
\end{equation}
so that
\begin{equation}
 \delta_R=\frac12\arg S,\qquad
 \delta_I=-\frac12\ln|S|.
 \label{eq:uncoupledparts}
\end{equation}
The phase is chosen continuously from the threshold.

For coupled triplet channels, we use a Stapp-type parametrization in the $LSJ$-basis phase convention employed in our LO analysis~\cite{Stapp:1956mz,Xiao:2024nbarN},
\begin{equation}
 S=\begin{pmatrix}
 e^{2i\delta_-}\cos2\epsilon_J &
 -i e^{i(\delta_-+\delta_+)}\sin2\epsilon_J\\
 -i e^{i(\delta_-+\delta_+)}\sin2\epsilon_J &
 e^{2i\delta_+}\cos2\epsilon_J
 \end{pmatrix},
 \label{eq:stapp}
\end{equation}
where $\delta_-=\delta_{J-1}$ and $\delta_+=\delta_{J+1}$.  The minus sign in the off-diagonal elements follows from this $LSJ$-basis phase choice.  Relative to the alternative convention with $+i$ in the off-diagonal elements, Eq.~\eqref{eq:stapp} corresponds to $\epsilon_J\to-\epsilon_J$.  In the presence of annihilation, both phase shifts and the mixing angle $\epsilon_J$ are complex.

\section{Numerical setup}
\label{sec:numerical}

We use the isospin-averaged masses $m_N=939$ MeV and $m_\pi=138$ MeV, together with $f_\pi=92.4$ MeV and the effective axial coupling $g_A=1.29$, chosen to account for the Goldberger--Treiman discrepancy in the OPE strength~\cite{Machleidt:2011zz,Baru:2011bw}.  For the subleading pion--nucleon LECs, we adopt the EOMS $\pi N$  values of Ref.~\cite{Chen:2012nx},
\begin{equation}
 (c_1,c_2,c_3,c_4)=(-1.39,\,4.01,\,-6.61,\,3.92)~\mathrm{GeV}^{-1}.
 \label{eq:numerical-inputs}
\end{equation}
The regulator cutoff is fixed at $\Lambda=700$ MeV.  For the conditional uncertainty estimate below, we take the same value for the breakdown scale, $\Lambda_b=700$ MeV.

\subsection{Fitting strategy}
\label{subsec:fit-strategy}

At each order and for each isospin, the complex short-range LECs are fitted to the energy-dependent Nijmegen PWA~\cite{Zhou:2012ui} for all partial waves with $J\leq2$ and $p_{\rm lab}=100$--$600$ MeV/$c$.   We have also verified numerically that the elastic $S$ matrices satisfy  $S^\dagger S \leq 1$ over the displayed energy range.

Motivated by the effective chi-square diagnostic of Ref.~\cite{Epelbaum:2014efa}, we quantify the agreement with the PWA using the effective $\widetilde\chi^2$.  The Nijmegen $\bar NN$ PWA does not provide statistical uncertainties, so we define
\begin{align}
 \frac{\widetilde\chi^2}{N_{\rm data}}={}&
 \frac{1}{N_{\rm data}}\sum_{i}
 \Biggl\{
 \left[\frac{\operatorname{Re}\delta_i-
 \operatorname{Re}\delta_i^{\rm PWA}}{\Delta_0}\right]^2
 \nonumber\\
 &\quad+
 \left[\frac{\operatorname{Im}\delta_i-
 \operatorname{Im}\delta_i^{\rm PWA}}{\Delta_0}\right]^2
 \Biggr\}, \qquad \Delta_0=1^\circ ,
 \label{eq:effective-chi2}
\end{align}
where $N_{\rm data}$ refers to the number of data points fitted, $\delta_i$ denotes either a phase shift or a  mixing angle and $i$  labels the fitted data points. The common scale $\Delta_0$ only renders the $\widetilde\chi^2$ dimensionless; it is not an assigned data uncertainty. 

\subsection{Bayesian truncation uncertainty}
\label{sec:uncertainty-method}

The truncation bands are estimated with the Bayesian EFT framework of Refs.~\cite{Furnstahl:2015rha,Melendez:2017phj,Melendez:2019izc}, following its implementation for the covariant $NN$ interaction in Ref.~\cite{Lu:2021gsb}.  For an observable $X$ (the real or imaginary part of a phase shift or mixing angle), we write
\begin{align}
 X&=X_{\rm ref}\sum_{\substack{n=0\\ n\ne1}}^\infty c_nQ^n,
 \label{eq:bayes-expansion}\\
 Q&=\frac{\max(p_{\rm cm},m_\pi)}{\Lambda_b},\qquad
 p_{\rm cm}=\sqrt{\frac{m_NT_{\rm lab}}{2}},
 \label{eq:phase-Q}
\end{align}
where $T_{\rm lab}$ is the laboratory kinetic energy and $\Lambda_b=700$ MeV.  The $Q^1$ term is absent in the adopted chiral counting.  Denoting the LO, NLO, and NNLO predictions by $X^{(0)}$, $X^{(2)}$, and $X^{(3)}$, respectively, we define
\begin{align}
 D_0&=X^{(0)},\qquad D_2=X^{(2)}-X^{(0)},\nonumber\\
 D_3&=X^{(3)}-X^{(2)},\nonumber\\
 X_{\rm ref}^{(k)}&=
 \max_{n\in\mathcal K_k}\frac{|D_n|}{Q^n},\nonumber\\
 c_n^{(k)}&=\frac{D_n}{X_{\rm ref}^{(k)}Q^n},
 \label{eq:bayes-coefficients}
\end{align}
where $\mathcal K_2=\{0,2\}$ and $\mathcal K_3=\{0,2,3\}$.  

As in Ref.~\cite{Lu:2021gsb}, the coefficients are assigned independent zero-mean Gaussian priors with a common width $\bar c$, together with a log-uniform prior for $\bar c$,
\begin{align}
 p(c_n|\bar c)&=\frac{1}{\sqrt{2\pi}\bar c}
 \exp\!\left(-\frac{c_n^2}{2\bar c^2}\right),\nonumber\\
 p(\bar c)&\propto\frac{1}{\bar c},
 \qquad 0.5\leq\bar c\leq10.
 \label{eq:bayes-prior}
\end{align}
The coefficient fixed to unit magnitude by $X_{\rm ref}^{(k)}$ is excluded from the inference of $\bar c$.  Denoting its index by $m_k$,
\begin{align}
 p(\bar c|\bm c_k)&\propto
 \bar c^{-(N_c+1)}
 \exp\!\left[-\frac{C_k^2}{2\bar c^2}\right],\nonumber\\
 C_k^2&=\sum_{\substack{n\in\mathcal K_k\\n\ne m_k}}
 \left(c_n^{(k)}\right)^2,
 \label{eq:bayes-cbar-posterior}
\end{align}
where $N_c=1$ at NLO and $N_c=2$ at NNLO.  We approximate the omitted remainder by the next ten powers,
\begin{align}
 \delta X_k&=X_{\rm ref}^{(k)}
 \sum_{n=k+1}^{k+10}c_nQ^n,\nonumber\\
 \sigma_k^2(\bar c)&=
 \left(X_{\rm ref}^{(k)}\right)^2\bar c^2
 \sum_{n=k+1}^{k+10}Q^{2n},
 \label{eq:bayes-remainder}
\end{align}
so that the posterior for the omitted contribution is
\begin{align}
 p(\delta X_k|\bm c_k)&=
 \int_{0.5}^{10}d\bar c\,
 \frac{\exp[-\delta X_k^2/(2\sigma_k^2)]}
 {\sqrt{2\pi\sigma_k^2}}\,
 p(\bar c|\bm c_k),
 \label{eq:bayes-remainder-posterior}\\
 0.68&=\int_{-dX_k}^{dX_k}
 p(\delta X_k|\bm c_k)\,d(\delta X_k).
 \label{eq:bayes-68}
\end{align}
Here $\sigma_k=\sigma_k(\bar c)$ in the first line.  The real and imaginary parts are treated independently.  Since the LO input is the hybrid interaction of Ref.~\cite{Xiao:2024nbarN}, whereas NLO and NNLO use the unified complex-contact interaction introduced here, the resulting 68\% degree-of-belief (DoB) intervals should be regarded as conditional EFT truncation estimates.

\section{Results and discussion}
\label{sec:scattering-results}
\subsection{Phase shifts}
\label{sec:phase shifts}

Table~\ref{tab:phase-chi2} gives the effective $\widetilde\chi^2/N_{\rm data}$ of Eq.~\eqref{eq:effective-chi2}. The totals show a substantial overall improvement at NNLO, although the reduction is not uniform in every phase parameter.  The fitted LEC values are listed in Table~\ref{tab:lecs}.

\begin{table*}[htb]
\caption{Effective $\widetilde\chi^2/N_{\rm data}$ at NLO and NNLO.}
\label{tab:phase-chi2}
\begin{ruledtabular}
\begin{tabular}{lcc@{\qquad}lcc}
$I=0$  & NLO & NNLO &
$I=1$  & NLO & NNLO \\
\hline
Total               & 0.124 & 0.027 & Total               & 0.044 & 0.010 \\
${}^{11}S_0$        & 0.065 & 0.011 & ${}^{31}S_0$        & 0.014 & 0.023 \\
${}^{13}P_0$        & 0.183 & 0.029 & ${}^{33}P_0$        & 0.003 & 0.006 \\
${}^{11}P_1$        & 0.041 & 0.044 & ${}^{31}P_1$        & 0.011 & 0.008 \\
${}^{13}P_1$        & 0.087 & 0.045 & ${}^{33}P_1$        & 0.008 & 0.003 \\
${}^{13}S_1$        & 0.071 & 0.051 & ${}^{33}S_1$        & 0.047 & 0.020 \\
${}^{13}D_1$        & 0.052 & 0.027 & ${}^{33}D_1$        & 0.013 & 0.012 \\
${}^{1}\epsilon_1$  & 0.104 & 0.022 & ${}^{3}\epsilon_1$  & 0.001 & 0.005 \\
${}^{11}D_2$        & 0.096 & 0.055 & ${}^{31}D_2$        & 0.119 & 0.009 \\
${}^{13}D_2$        & 0.083 & 0.005 & ${}^{33}D_2$        & 0.206 & 0.026 \\
${}^{13}P_2$        & 0.584 & 0.025 & ${}^{33}P_2$        & 0.099 & 0.010 \\
${}^{13}F_2$        & 0.105 & 0.003 & ${}^{33}F_2$        & 0.003 & 0.000 \\
${}^{1}\epsilon_2$  & 0.019 & 0.003 & ${}^{3}\epsilon_2$  & 0.003 & 0.001 \\
\end{tabular}
\end{ruledtabular}
\end{table*}

\begin{table*}[htb]
\caption{Fitted complex LECs at NLO and NNLO, in units of $10^4~\mathrm{GeV}^{-2}$.}
\label{tab:lecs}
\centering
\scriptsize
\setlength{\tabcolsep}{3.4pt}
\begin{ruledtabular}
\begin{tabular}{c rr rr rr rr}
& \multicolumn{2}{c}{NLO, $I=0$} & \multicolumn{2}{c}{NLO, $I=1$}
& \multicolumn{2}{c}{NNLO, $I=0$} & \multicolumn{2}{c}{NNLO, $I=1$} \\
Coefficient & Re & Im & Re & Im & Re & Im & Re & Im \\
\hline
$C_1$ & 0.887 & 4.652 & -1.692 & 1.198 & -0.458 & 0.255 & 17.860 & -1.546 \\
$C_2$ & -1.218 & -7.048 & 1.382 & -2.373 & 0.477 & -0.194 & -7.492 & -4.891 \\
$C_3$ & 1.124 & 3.555 & -0.200 & 2.775 & 0.853 & 0.487 & 6.057 & 3.527 \\
$C_4$ & 1.631 & 3.031 & -0.238 & 1.191 & 0.453 & 0.250 & -0.778 & 3.107 \\
$C_5$ & 0.239 & -2.175 & -3.901 & 8.279 & 0.019 & 0.179 & 5.605 & 5.303 \\
$C_6$ & -1.660 & -5.347 & -0.169 & -0.754 & -0.386 & -0.348 & -4.239 & -2.928 \\
$C_7$ & -0.471 & 0.475 & 0.511 & -1.540 & -0.494 & -0.175 & 2.162 & -1.296 \\
$C_8$ & 1.142 & -0.590 & 1.364 & -1.134 & 0.857 & 0.089 & -7.534 & 1.686 \\
$C_9$ & -3.281 & -4.330 & -5.212 & 4.169 & -2.369 & -0.574 & 8.099 & 0.830 \\
$C_{10}$ & -1.176 & -8.696 & -1.446 & 2.170 & -3.703 & -0.695 & -0.614 & 2.961 \\
$C_{11}$ & 4.381 & -8.084 & 2.379 & -2.960 & 2.877 & 0.419 & -3.964 & -0.682 \\
$C_{12}$ & -2.324 & -9.368 & -1.926 & 2.934 & -1.411 & -0.528 & 1.891 & 1.327 \\
$C_{13}$ & -0.704 & 6.579 & 0.764 & -0.736 & -0.177 & 0.056 & 0.169 & -0.863 \\
$C_{14}$ & -0.377 & 0.569 & -2.916 & 3.558 & -0.941 & -0.081 & 16.990 & -1.294 \\
$C_{15}$ & 1.776 & -5.137 & 0.712 & 0.019 & 0.400 & -0.023 & 3.155 & -2.520 \\
$C_{16}$ & 2.653 & 0.813 & -0.297 & 0.451 & 0.231 & 0.240 & 5.454 & 1.503 \\
$C_{17}$ & -0.801 & 2.268 & -0.421 & 0.867 & 0.161 & 0.097 & -2.340 & 1.491 \\
$D_{{}^1D_2}$ & -1.371 & 2.792 & -0.840 & 0.728 & -0.693 & -0.298 & 0.654 & 0.651 \\
$D_{{}^3D_2}$ & -3.189 & -0.795 & -3.493 & 3.659 & -2.685 & -0.618 & 6.018 & 3.176 \\
$D_{{}^3P_2}$ & -0.069 & 1.436 & 0.320 & -0.234 & -1.130 & 0.323 & -0.779 & 0.741 \\
$D_{{}^3P_2-{}^3F_2}$ & -3.226 & 7.560 & -1.723 & 2.122 & -1.263 & -0.182 & 1.078 & 2.457 \\
$D_{{}^3F_2}$ & -1.781 & 0.002 & -1.824 & -0.951 & 0.932 & -0.345 & -1.750 & -0.803 \\
\end{tabular}
\end{ruledtabular}
\end{table*}

Figs.~\ref{fig:covariant-I0-uncoupled}--\ref{fig:covariant-I1-coupled} show the real and imaginary parts of the phase shifts and mixing angles.  The covariant NLO and NNLO results are displayed with the 68\% Bayesian DoB bands defined above and are compared with the heavy-baryon (HB) NNLO and N$^3$LO results of Ref.~\cite{Dai:2017ont}.  The covariant results are evaluated at the discrete momenta $p_{\rm lab}=100,200,\ldots,800$ MeV/$c$ and connected by lines; the corresponding curves are therefore less smooth than the HB ones.

\subsubsection{Isoscalar partial waves}
\label{sec:I0-results}

\begin{figure*}[htpb]
  \centering
  \includegraphics[width=\textwidth]{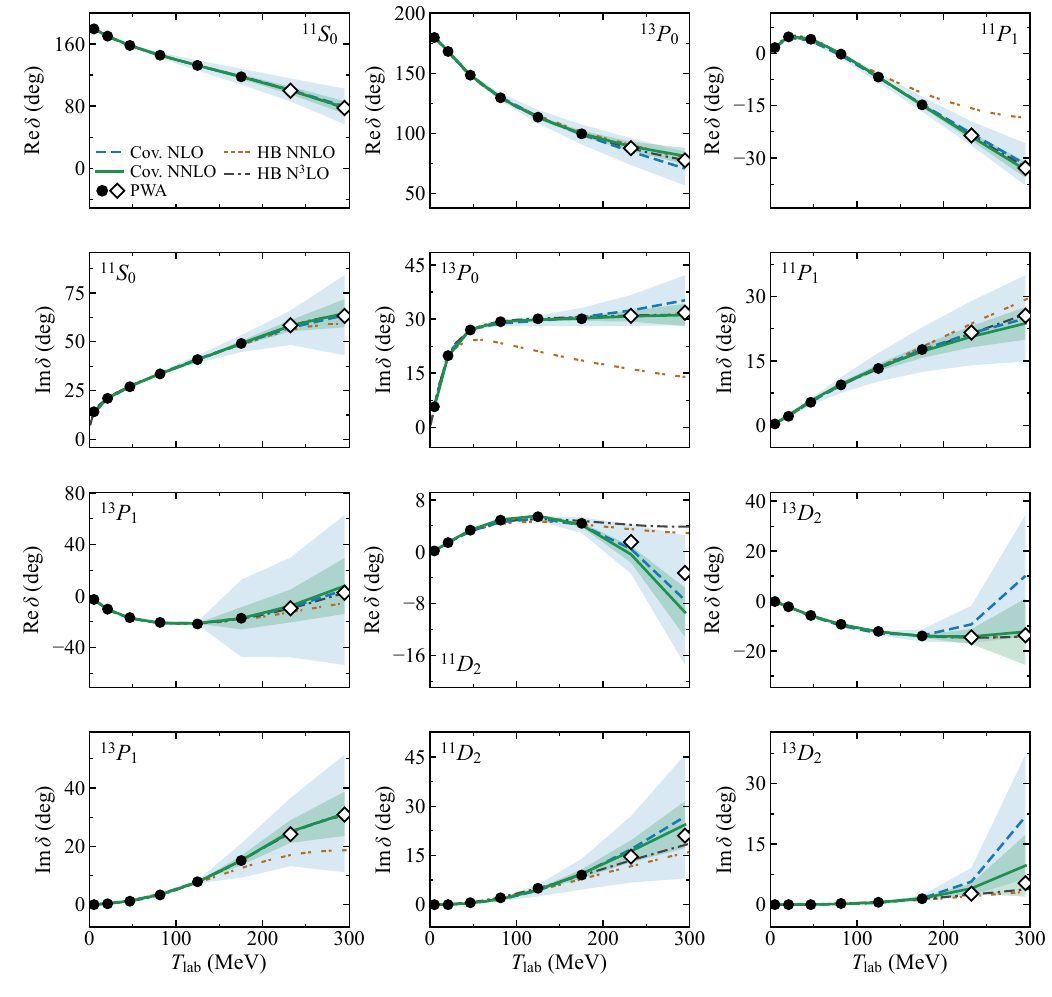}
  \caption{Real and imaginary parts of the $I=0$ uncoupled-channel phase   shifts.  Blue dashed and green solid curves show the covariant NLO and NNLO   results at $\Lambda=700$ MeV, with shaded 68\% Bayesian DoB bands.  Orange dotted and gray dash-dotted curves show the HB NNLO and N$^3$LO results of   Ref.~\cite{Dai:2017ont} for $R=0.9$ fm.  Black circles denote Nijmegen PWA   values included in the fit, and open diamonds denote the higher-energy comparison points.}
  \label{fig:covariant-I0-uncoupled}
\end{figure*}

\begin{figure*}[htpb]
  \centering
  \includegraphics[width=\textwidth]{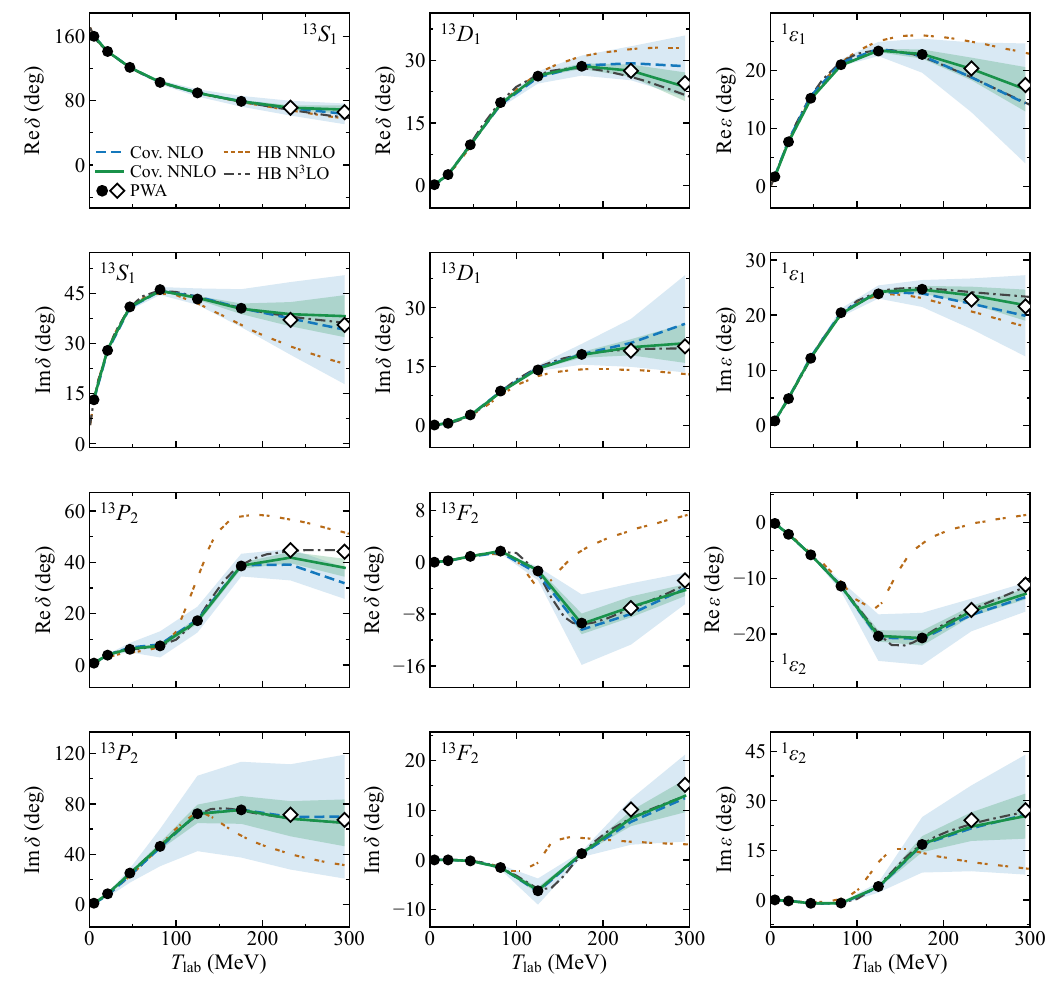}
  \caption{Same as Fig.~\ref{fig:covariant-I0-uncoupled}, but for the   $I=0$ coupled-channel phase shifts and mixing angles.}
  \label{fig:covariant-I0-coupled}
\end{figure*}

The isoscalar results are shown in Figs.~\ref{fig:covariant-I0-uncoupled} and~\ref{fig:covariant-I0-coupled}. Both chiral orders describe the overall energy dependence of most low partial waves over the fitted region, but the NNLO corrections vary significantly among channels. The real \({}^{11}S_0\) phase shift is nearly unchanged, while its imaginary part moves closer to the PWA. A more pronounced improvement occurs in \({}^{13}P_0\), both components follow the PWA more closely at NNLO, and the imaginary part remains approximately saturated above the fitted region. In contrast, the HB NNLO imaginary \({}^{13}P_0\) phase shifts decrease with energy, whereas the HB N$^3$LO and covariant NNLO curves reproduce the qualitative PWA trend. The agreement is less uniform for the \(^{11}P_1\) and \(^{13}P_1\) waves. For \(^{11}P_1\), the NNLO real part of the phase shift follows the decreasing PWA trend but becomes slightly too negative at higher energies, while its imaginary part falls below the PWA. The corresponding fit quality is marginally worse than at NLO. In the \(^{13}P_1\) channel, the real part of the phase shift increases more rapidly than the PWA beyond the fitted region, whereas the imaginary part remains reasonably close to the PWA values.

The two uncoupled \(D\) waves exhibit rather different order-by-order behavior. In \({}^{13}D_2\), the NLO real and imaginary parts rise rapidly beyond the fit region, and the real part of the phase shifts even become positive at the highest plotted energy, contrary to the PWA. NNLO preserves the negative real part of the phase shifts and considerably reduces the rise of the imaginary part, giving better agreement with both the PWA and the HB predictions. For \(^{11}D_2\), however, neither order describes both components satisfactorily at the two higher-energy comparison points. The real part of the phase shifts become too negative, while the imaginary parts are overestimated. In particular, the two high-energy PWA real-phase values are outside the conditional NNLO band. The isoscalar \(D\) waves therefore do not exhibit uniformly improved extrapolation.

In the coupled \(^{13}S_1\)–\(^{13}D_1\) system, the \(^{13}S_1\) phase shifts change little within the fitted region. The principal NNLO changes occur in \({}^{13}D_1\) and \({}^{1}\epsilon_1\). Their real parts decrease beyond the fitted region, broadly following the PWA, and the imaginary parts are also described more closely. The \({}^{13}P_2\)–\({}^{13}F_2\) system shows a substantial improvement within the fit region. The NNLO description, however, remains imperfect beyond the fitted region. The real \(^{13}P_2\) phase shift is underestimated at both higher-energy comparison points, whereas its imaginary part remains close to the PWA. For \(^{13}F_2\), the NNLO real part of the phase shift reproduces the negative trend of the PWA, while its imaginary part is slightly underestimated at the highest energy. The real mixing angle \(^{1}\epsilon_2\) follows the pronounced decrease and subsequent recovery seen in the PWA, whereas its imaginary part lies slightly below the PWA at the two prediction points. Thus, the NNLO and HB N$^3$LO descriptions are comparable in several fitted components, while the HB N$^3$LO result is closer to the PWA for some extrapolated quantities.

NLO and NNLO employ the same short-range operator basis and the same number of complex LECs, with independent fits at the two orders.  The improvement cannot therefore be attributed to additional NNLO contact operators.  The new contribution is the real, $c_i$-dependent $Q^3$ TPE, which changes the momentum and tensor dependence of the intermediate-range interaction. Together with the refit of the same complex contact terms and the nonperturbative solution of the Kadyshevsky equation, this also changes the imaginary phase parameters.  

\subsubsection{Isovector partial waves}
\label{sec:I1-results}

\begin{figure*}[htpb]
  \centering
  \includegraphics[width=\textwidth]{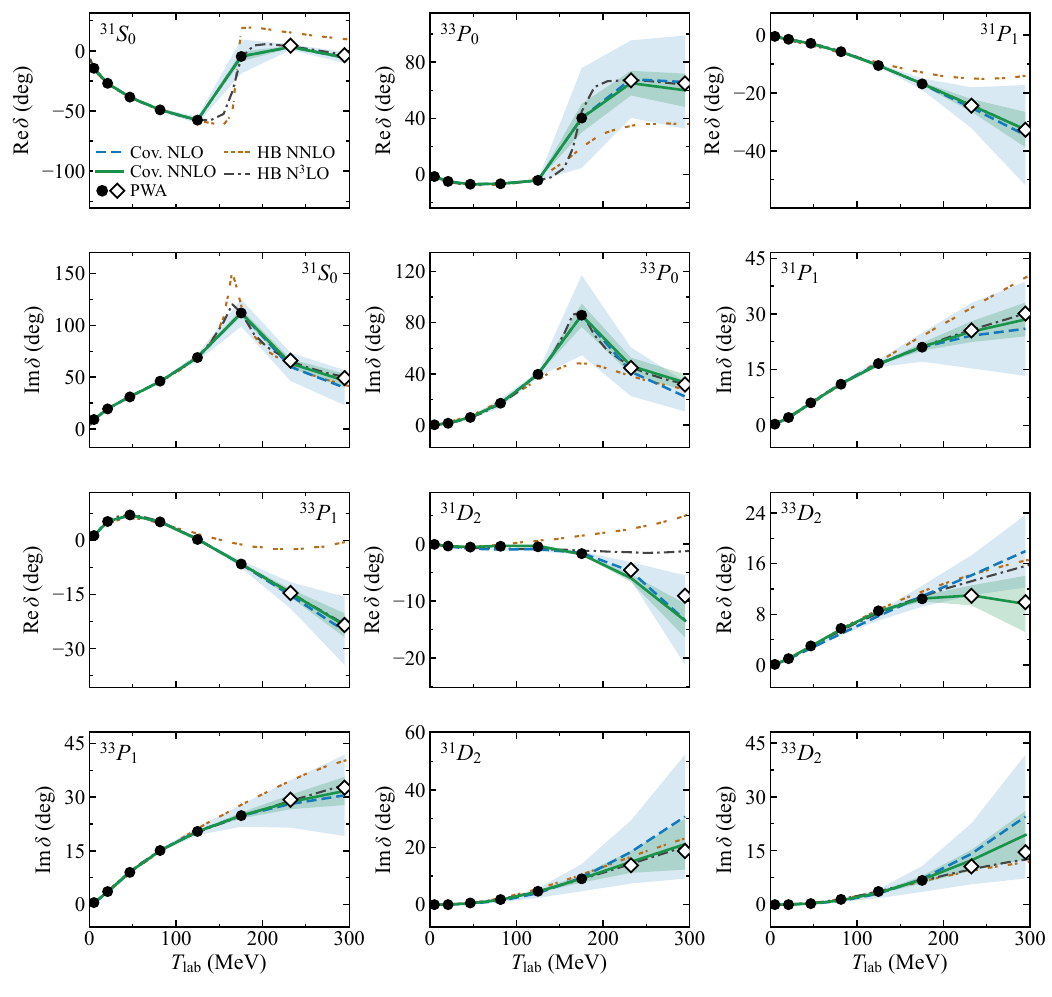}
  \caption{Same as Fig.~\ref{fig:covariant-I0-uncoupled}, but for the  $I=1$ uncoupled channels.}
  \label{fig:covariant-I1-uncoupled}
\end{figure*}

\begin{figure*}[htpb]
  \centering
  \includegraphics[width=\textwidth]{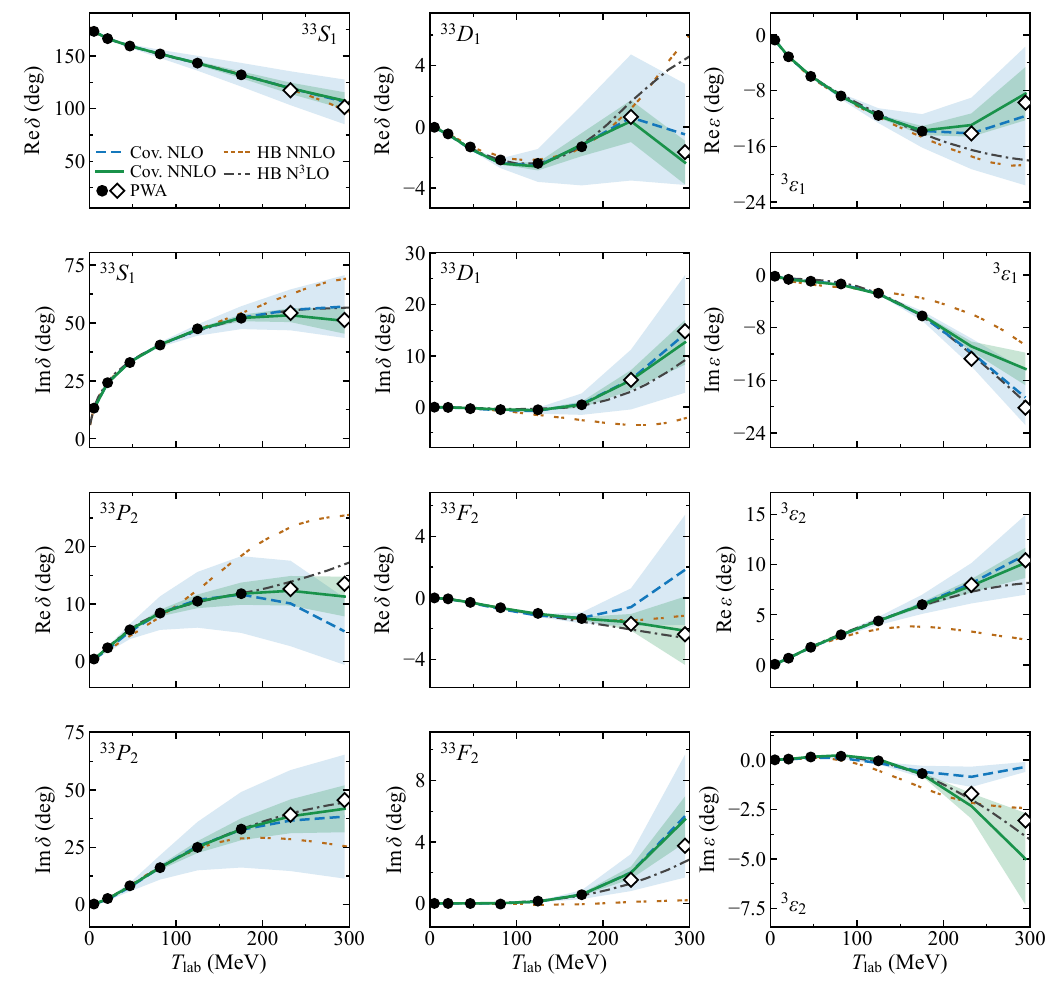}
  \caption{Same as Fig.~\ref{fig:covariant-I0-uncoupled}, but for the  $I=1$ coupled-channel phase shifts and mixing angles.}
  \label{fig:covariant-I1-coupled}
\end{figure*}

The isovector results are displayed in Figs.~\ref{fig:covariant-I1-uncoupled} and \ref{fig:covariant-I1-coupled}.  The strongly energy-dependent \({}^{31}S_0\) and \({}^{33}P_0\) phase shifts are broadly reproduced by both covariant orders in the fit region. In these channels, the real part of the phase shift changes rapidly toward the upper end of the fitted interval, while the imaginary part reaches a maximum and subsequently decreases. The HB N$^3$LO interaction also reproduces these resonance-like PWA trends. At the comparison points, the NNLO imaginary part of the phase shifts remains close to the PWA, but the fit measure for each of these two channels is slightly larger at NNLO than at NLO. At the highest energy, the NNLO real \(^{33}P_0\) phase shift does not show a clear improvement over NLO. The smoother \(^{31}P_1\) and \(^{33}P_1\) phase shifts exhibit only modest order-by-order changes within the fitted region. At higher energies, NNLO generally provides a slightly better description of both the real and imaginary components, although the differences between the two orders remain small.

For the uncoupled isovector \(D\) waves, NNLO markedly improves the fitted energy dependence, as quantified by the reductions in Table~\ref{tab:phase-chi2}. The improvement for \({}^{33}D_2\) largely persists at the two higher-energy points; the NLO real part of the phase shift continues to increase whereas the PWA turns over, and the NNLO result reproduces that turnover. Its imaginary part also moves closer to the PWA, although it remains too large at the highest point. The extrapolation in \({}^{31}D_2\) is more mixed. NNLO substantially improves the imaginary part of the phase shift, but the real part decreases too quickly beyond the fitted interval. Improvements within the fitted region therefore need not carry over to all extrapolated components.

The \({}^{33}S_1\)–\({}^{33}D_1\) system shows relatively modest changes in the fit region. Both orders reproduce the dominant \({}^{33}S_1\) phase shift, with a better high-energy imaginary part at NNLO. The real \(^{33}D_1\) phase shifts initially decrease and subsequently turn upward, with both covariant orders reproducing this behavior. At the highest energy, NNLO better describes the downward turn observed in the PWA. The imaginary part remains small at low energies and increases rapidly at higher energies. Both orders reproduce this trend, although NLO is slightly closer to the PWA at the highest energy.   The mixing angle \(^{3}\epsilon_1\) exhibits different order-by-order behavior in its two components.  The NNLO real part follows the upward turn of the PWA at higher energies and is closer to the highest-energy comparison point. In contrast, the magnitude of the imaginary part is underestimated at NNLO, while NLO remains closer to the PWA at that energy.

In the coupled \({}^{33}P_2\)–\({}^{33}F_2\) system, the two orders separate appreciably beyond the fit region. The NLO real \({}^{33}P_2\) phase shifts bend downward, the real \({}^{33}F_2\) phase shifts turn positive, and the imaginary \({}^{3}\epsilon_2\) returns toward zero. All three  features disagree with the PWA trends. These deviations may reflect the limited accuracy of the NLO extrapolation beyond the fitted energy range. Higher-order corrections, together with the associated refit of complex contact couplings, can modify the balance between the diagonal interactions and $P$--$F$ mixing, thereby changing the energy dependence of all three quantities. At NNLO, the real \({}^{33}P_2\) phase shifts remain closer to the observed plateau, the real \({}^{33}F_2\) phase shifts stay negative, and the imaginary \({}^{3}\epsilon_2\) bend downward. The latter becomes too negative at the highest point, while the NNLO imaginary \({}^{33}F_2\) phase shifts rise above the PWA. The real \({}^{3}\epsilon_2\) remain well described throughout the plotted interval. In particular, the separation of the NLO and NNLO bands in the high-energy imaginary \({}^{3}\epsilon_2\) indicates appreciable order dependence and does not demonstrate  convergence. The NNLO description is comparable to HB N$^3$LO for several components, but HB N$^3$LO performs better for some extrapolated imaginary quantities.

As for \(I=0\), the \(I=1\) NLO and NNLO calculations use the same 22 complex contact structures with independently fitted coefficients. The isospin dependence of the \(Q^3\) TPE contributions provides different dispersive corrections in the two sectors. Through the refitting and iteration of the complex interaction, these corrections also affect absorptive observables. In particular, changes in the imaginary \({}^{3}\epsilon_2\) should not be interpreted as an additional, explicitly calculated NNLO annihilation mechanism.

\subsection{Pole analysis}
\label{sec:poles}

Our previous LO study~\cite{Xiao:2024nbarN} identified pole candidates in the ${}^{11}S_0$, ${}^{13}P_0$, ${}^{13}S_1$, and ${}^{33}S_1$ channels.  We now revisit these structures with the NLO and NNLO interactions.  The pole energy is written as
\begin{equation}
 E_{\rm pole}=W_p-2m_N=E_R-i\frac{\Gamma}{2},
 \label{eq:pole-definition}
\end{equation}
where $W_p$ is the complex c.m. energy, $E_R$ is measured relative to the $\bar NN$ threshold, and $\Gamma=-2\,\mathrm{Im}\,E_{\rm pole}$.  The scattering equation is analytically continued to complex energy, and a pole
is located when the discretized operator
\begin{equation}
 A(W)=\unit-V(W)G_0(W)
 \label{eq:pole-condition}
\end{equation}
acquires a zero mode.  Both the interaction and the free propagator $G_0$ are evaluated at the complex energy. 

The pole uncertainties are obtained from the same Bayesian model, applied separately to $E_R$ and $\Gamma$.  Eqs.~\eqref{eq:bayes-coefficients}--\eqref{eq:bayes-68} are used with the order-dependent expansion parameter
\begin{align}
 Q_k^{\rm pole}&=\frac{\max\!\left(m_\pi,\{|q_n|\}_{n\in\mathcal K_k}\right)}
 {\Lambda_b},\nonumber\\
 q_n^2&=\frac{(2m_N+E_{\rm pole}^{(n)})^2}{4}-m_N^2,
 \label{eq:pole-Q}
\end{align}
where $\mathcal K_2=\{0,2\}$ and $\mathcal K_3=\{0,2,3\}$.   

\begin{table*}[htb]
\centering
\caption{Pole energies relative to the $\bar NN$ threshold, $E_{\rm pole}=E_R-i\Gamma/2$ (MeV). Uncertainties in $E_R$ and $\Gamma$ are 68\% conditional Bayesian DoB intervals. HB ranges reflect cutoff variation~\cite{Kang:2013uia}. Dashes indicate that no pole candidate is reported.}
\label{tab:pole-positions}
\small
\begin{ruledtabular}
\begin{tabular}{lllcc}
Partial wave & Approach & Order & $E_R$ & $\Gamma$ \\
\hline
${}^{11}S_0$
 & This work & LO
 & $-179.371$ & $277.498$ \\
 & This work & NLO
 & \multicolumn{2}{c}{ --- } \\
 & This work & NNLO
 &\multicolumn{2}{c}{ --- } \\
 & HB & NLO
 & \multicolumn{2}{c}{ --- } \\
 & HB & NNLO
 & \multicolumn{2}{c}{ --- } \\
\hline
${}^{13}P_0$
 & This work & LO
 & $-1.835$ & $41.965$ \\
 & This work & NLO
 & $-1.702\pm0.034$ & $36.450\pm1.317$ \\
 & This work & NNLO
 & $-2.632\pm0.156$ & $44.184\pm1.304$ \\
 & HB & NLO
 & $[-1.1,\,+1.9]$ & $[35.6,\,44.8]$ \\
 & HB & NNLO
 & $[-3.7,\,-0.2]$ & $[44.0,\,52.8]$ \\
\hline
${}^{13}S_1$--${}^{13}D_1$
 & This work & LO
 & $+24.396$ & $102.320$ \\
 & This work & NLO
 & $-6.523\pm13.552$ & $136.783\pm15.977$ \\
 & This work & NNLO
 & $+0.274\pm3.812$ & $132.537\pm4.129$ \\
 & HB & NLO
 & $[5.6,\,7.7]$ & $[98.4,\,121.0]$ \\
 & HB & NNLO
 & $[4.8,\,21.3]$ & $[121.2,\,149.8]$ \\
\hline
${}^{33}S_1$--${}^{33}D_1$
 & This work & LO
 & $-30.958$ & $316.506$ \\
 & This work & NLO
 & \multicolumn{2}{c}{ --- } \\
 & This work & NNLO
 & \multicolumn{2}{c}{ --- }\\
 & HB & NLO
 & \multicolumn{2}{c}{ --- }\\
 & HB & NNLO
 & \multicolumn{2}{c}{ --- } \\
\end{tabular}
\end{ruledtabular}
\end{table*}

Table~\ref{tab:pole-positions} summarizes the pole candidates identified in this work. The ${}^{13}P_0$ candidate exhibits comparatively weak order dependence: its real part remains a few MeV below the $\bar N N$ threshold, while its width stays close to 40 MeV at NLO and NNLO. These results suggest a persistent near-threshold pole candidate within the interactions considered, although its dependence on the regulator and short-range parameterization remains to be investigated.

The $S$-wave results exhibit channel-dependent behavior. Within the present analysis, no stable poles are identified in the ${}^{11}S_0$ and ${}^{33}S_1$--${}^{33}D_1$ channels at either NLO or NNLO. In the ${}^{13}S_1$--${}^{13}D_1$ channel, pole candidates are found at both orders, with their real parts close to the $\bar N N$ threshold. The difference between the LO and NLO positions is substantial, whereas the NLO and NNLO results are considerably closer. This smaller variation suggests reduced order dependence, although the physical interpretation of this candidate, particularly its association with a specific observed structure, still warrants caution.

At NLO and NNLO, both the present calculation and the HB study of Ref.~\cite{Kang:2013uia} identify pole candidates in the ${}^{13}P_0$ and ${}^{13}S_1$--${}^{13}D_1$ channels, although their predicted positions differ. For ${}^{13}P_0$, the NNLO central values of $E_R$ and $\Gamma$ lie within the respective HB cutoff ranges. At NLO, the width also falls within the HB range, whereas the real part lies below it. In the ${}^{13}S_1$--${}^{13}D_1$ channel, the present central values of $E_R$ are lower than the HB ranges at both orders. The NLO width lies above the HB range, while the NNLO width falls within it. The absence of stable candidates in the ${}^{11}S_0$ and ${}^{33}S_1$--${}^{33}D_1$ channels at NLO and NNLO is qualitatively consistent with the absence of near-threshold poles reported in that study. The HB ranges reflect cutoff variation and should be distinguished from the conditional Bayesian truncation intervals quoted here.

To complement the pole analysis, we determine the complex threshold parameters of the same channels through the generalized effective-range expansion,
\begin{equation}
 k^{2\ell+1}\cot\delta_\ell(k)
 =-\frac{1}{a_\ell}
 +\frac{1}{2}r_\ell k^2
 +\mathcal O(k^4),
 \label{eq:effective-range-expansion}
\end{equation}
where $k$ is the on-shell c.m. momentum. For $S$ waves ($\ell=0$), $a_0$ and $r_0$ are the scattering length and effective range, both in units of fm. For ${}^{13}P_0$ ($\ell=1$), $a_1$ is the scattering volume in $\mathrm{fm}^3$, while $r_1$ is the effective-range parameter in $\mathrm{fm}^{-1}$. These parameters are generally complex because annihilation channels remain open above the $\bar NN$ threshold.

The uncertainties of the threshold parameters are estimated following the same procedure as for the pole energies. We apply these equations separately to the real and imaginary parts of $a_\ell$ and $r_\ell$, using their LO, NLO, and NNLO values and the threshold expansion parameter $Q=m_\pi/\Lambda_b$. The priors and the prescription for the conditional 68\% DoB intervals remain unchanged.

Table~\ref{tab:threshold-parameters} shows that the complex scattering lengths and the ${}^{13}P_0$ scattering volume are close to the corresponding HB predictions of Ref.~\cite{Dai:2017ont}. The similarity in both real and imaginary parts indicates comparable leading threshold behavior for elastic scattering and annihilation. This agreement is consistent with the common PWA constraints, despite differences in the formulation and short-range parameterization of the interactions. We also report the complex effective-range parameters for all four channels, which are not provided in that reference.

The threshold parameters complement the pole analysis by constraining the amplitudes on the physical axis near threshold. Their agreement with HB does not require identical pole positions, which depend on the continuation of the amplitudes into the complex-energy plane.

\begin{table*}[htb]
\centering
\caption{Complex threshold parameters compared with HB results at $R=0.9$ fm~\cite{Dai:2017ont}. Quoted uncertainties are 68\% conditional Bayesian DoB intervals for the real and imaginary parts. Units are fm for $a_0,r_0$, $\mathrm{fm}^3$ for $a_1$, and $\mathrm{fm}^{-1}$ for $r_1$. Dashes denote values not reported in the HB reference.}
\label{tab:threshold-parameters}
\small
\setlength{\tabcolsep}{3pt}
\begin{ruledtabular}
\begin{tabular}{lllrrrr}
Partial wave & Approach & Order & $\operatorname{Re}a_\ell$ & $\operatorname{Im}a_\ell$ & $\operatorname{Re}r_\ell$ & $\operatorname{Im}r_\ell$ \\
\hline
${}^{11}S_0$ & This work & LO & $-0.245$ & $-1.197$ & $-3.804$ & $-11.834$ \\
 & This work & NLO & $-0.206\pm0.009$ & $-1.172\pm0.012$ & $-4.584\pm0.178$ & $-11.641\pm0.111$ \\
 & This work & NNLO & $-0.212\pm0.001$ & $-1.215\pm0.007$ & $-3.998\pm0.095$ & $-11.255\pm0.063$ \\
 & HB & NLO & $-0.21$ & $-1.20$ & --- & --- \\
 & HB & NNLO & $-0.21$ & $-1.22$ & --- & --- \\
 & HB & N$^3$LO & $-0.20$ & $-1.23$ & --- & --- \\
\hline
${}^{13}P_0$ & This work & LO & $-3.315$ & $-7.820$ & $-1.126$ & $-1.773$ \\
 & This work & NLO & $-3.367\pm0.031$ & $-8.296\pm0.126$ & $-1.129\pm0.010$ & $-1.673\pm0.027$ \\
 & This work & NNLO & $-2.998\pm0.059$ & $-7.684\pm0.099$ & $-1.212\pm0.013$ & $-1.803\pm0.021$ \\
 & HB & NLO & $-3.62$ & $-8.05$ & --- & --- \\
 & HB & NNLO & $-3.18$ & $-8.02$ & --- & --- \\
 & HB & N$^3$LO & $-2.83$ & $-7.82$ & --- & --- \\
\hline
${}^{13}S_1$--${}^{13}D_1$ & This work & LO & $1.299$ & $-0.930$ & $1.295$ & $-1.248$ \\
 & This work & NLO & $1.388\pm0.023$ & $-0.873\pm0.015$ & $1.175\pm0.029$ & $-1.225\pm0.012$ \\
 & This work & NNLO & $1.401\pm0.004$ & $-0.898\pm0.004$ & $1.161\pm0.004$ & $-1.216\pm0.002$ \\
 & HB & NLO & $1.33$ & $-0.85$ & --- & --- \\
 & HB & NNLO & $1.39$ & $-0.89$ & --- & --- \\
 & HB & N$^3$LO & $1.42$ & $-0.88$ & --- & --- \\
\hline
${}^{33}S_1$--${}^{33}D_1$ & This work & LO & $0.398$ & $-0.849$ & $-1.372$ & $-0.488$ \\
 & This work & NLO & $0.441\pm0.010$ & $-0.928\pm0.019$ & $-1.032\pm0.077$ & $-0.893\pm0.091$ \\
 & This work & NNLO & $0.439\pm0.001$ & $-0.933\pm0.003$ & $-1.087\pm0.013$ & $-0.971\pm0.016$ \\
 & HB & NLO & $0.44$ & $-0.92$ & --- & --- \\
 & HB & NNLO & $0.45$ & $-0.95$ & --- & --- \\
 & HB & N$^3$LO & $0.44$ & $-0.96$ & --- & --- \\
\end{tabular}
\end{ruledtabular}
\end{table*}

\section{Summary and outlook}
\label{sec:summary}

We have constructed covariant chiral $\bar NN$ interactions at NLO and NNLO. The long-range part is obtained from the covariant $NN$ pion-exchange interaction by $G$-parity, while short-range elastic and annihilation effects are described by a common set of complex covariant contact terms.  The LECs are fitted to the Nijmegen PWA for partial waves with $J\leq2$, and the scattering amplitudes are obtained from the Kadyshevsky equation with $\Lambda=700$ MeV.

Both NLO and NNLO provide a quantitative description of the low-energy phase shifts and mixing angles.  The NNLO corrections are most relevant for several imaginary components and $J=2$ channels, while the overall NNLO quality is comparable to the HB N$^3$LO interaction.  Using the Bayesian procedure, we quote 68\% DoB intervals for the order-by-order truncation uncertainty.

The pole analysis shows a clearer hierarchy of stability.  The ${}^{13}P_0$ pole remains close to threshold with a similar width at NLO and NNLO and is therefore the most robust near-threshold candidate found here. The $S$-wave pole candidates show less uniform behavior. No stable poles are identified in the \(^{11}S_0\) and \(^{33}S_1\)–\(^{33}D_1\) channels at NLO or NNLO, while the \(^{13}S_1\)–\(^{13}D_1\) candidate exhibits appreciable order dependence, particularly relative to LO. Their physical interpretation therefore requires further investigation. We have also calculated the complex $S$-wave scattering lengths and effective ranges, together with the ${}^{13}P_0$ scattering volume and effective-range parameter. The predicted scattering lengths and scattering volume are close to the corresponding HB results. Future work should address regulator dependence and apply the higher-order $\bar NN$ amplitudes directly to measured near-threshold line shapes.

\section*{Acknowledgments}
This work is partly supported by the National Natural Science Foundation of China under Grant Nos.~12435007 and 12505096, the China Postdoctoral Science Foundation under Grant No.~2022M720360, and the Institute for Basic Science (IBS-R031-D1).

\bibliography{references}

@article{Gasser:1987rb,
    author = "Gasser, J. and Sainio, M. E. and Svarc, A.",
    title = "{Nucleons with chiral loops}",
    reportNumber = "BUTP-87-17",
    doi = "10.1016/0550-3213(88)90108-3",
    journal = "Nucl. Phys. B",
    volume = "307",
    pages = "779--853",
    year = "1988"
}

@article{Epelbaum:2014efa,
    author = "Epelbaum, E. and Krebs, H. and Mei{\ss}ner, U. G.",
    title = "{Improved chiral nucleon-nucleon potential up to next-to-next-to-next-to-leading order}",
    eprint = "1412.0142",
    archivePrefix = "arXiv",
    primaryClass = "nucl-th",
    doi = "10.1140/epja/i2015-15053-8",
    journal = "Eur. Phys. J. A",
    volume = "51",
    number = "5",
    pages = "53",
    year = "2015"
}

@article{Fettes:2000gb,
    author = "Fettes, Nadia and Mei{\ss}ner, Ulf-G. and Mojzis, Martin and Steininger, Sven",
    title = "{The Chiral Effective Pion-Nucleon Lagrangian of Order $p^4$}",
    eprint = "hep-ph/0001308",
    archivePrefix = "arXiv",
    reportNumber = "FZJ-IKP-TH-2000-04",
    doi = "10.1006/aphy.2000.6059",
    journal = "Annals Phys.",
    volume = "283",
    pages = "273--302",
    year = "2000",
    note = "Erratum: Annals Phys. 288, 249--250 (2001)"
}

@article{Weinberg:1991um,
    author = "Weinberg, Steven",
    title = "{Effective chiral Lagrangians for nucleon - pion interactions and nuclear forces}",
    reportNumber = "UTTG-03-91",
    doi = "10.1016/0550-3213(91)90231-L",
    journal = "Nucl. Phys. B",
    volume = "363",
    pages = "3--18",
    year = "1991"
}

@article{Weinberg:1990rz,
    author = "Weinberg, Steven",
    title = "{Nuclear forces from chiral Lagrangians}",
    reportNumber = "UTTG-31-90",
    doi = "10.1016/0370-2693(90)90938-3",
    journal = "Phys. Lett. B",
    volume = "251",
    pages = "288--292",
    year = "1990"
}

@article{Weinberg:1992yk,
    author = "Weinberg, Steven",
    title = "{Three body interactions among nucleons and pions}",
    eprint = "hep-ph/9209257",
    archivePrefix = "arXiv",
    reportNumber = "UTTG-11-92",
    doi = "10.1016/0370-2693(92)90099-P",
    journal = "Phys. Lett. B",
    volume = "295",
    pages = "114--121",
    year = "1992"
}

@article{Reinert:2017usi,
    author = "Reinert, P. and Krebs, H. and Epelbaum, E.",
    title = "{Semilocal momentum-space regularized chiral two-nucleon potentials up to fifth order}",
    eprint = "1711.08821",
    archivePrefix = "arXiv",
    primaryClass = "nucl-th",
    doi = "10.1140/epja/i2018-12516-4",
    journal = "Eur. Phys. J. A",
    volume = "54",
    number = "5",
    pages = "86",
    year = "2018"
}

@article{BES:2003aic,
    author = "Bai, J. Z. and others",
    collaboration = "BES",
    title = "{Observation of a near-threshold enhancement in the $p\bar p$ mass spectrum from radiative $J/\psi\to\gamma p\bar p$ decays}",
    eprint = "hep-ex/0303006",
    archivePrefix = "arXiv",
    reportNumber = "SLAC-PUB-9709, UH-511-1019-03",
    doi = "10.1103/PhysRevLett.91.022001",
    journal = "Phys. Rev. Lett.",
    volume = "91",
    pages = "022001",
    year = "2003"
}

@article{Belle:2002bro,
    author = "Abe, Kazuo and others",
    collaboration = "Belle",
    title = "{Observation of B+- ---\ensuremath{>} p anti-p K+-}",
    eprint = "hep-ex/0202017",
    archivePrefix = "arXiv",
    reportNumber = "KEK-PREPRINT-2001-170, BELLE-PREPRINT-2002-4",
    doi = "10.1103/PhysRevLett.88.181803",
    journal = "Phys. Rev. Lett.",
    volume = "88",
    pages = "181803",
    year = "2002"
}

@article{Belle:2002fay,
    author = "Abe, Kazuo and others",
    collaboration = "Belle",
    title = "{Observation of anti-B0 ---\ensuremath{>} D0(*) p anti-p}",
    eprint = "hep-ex/0205083",
    archivePrefix = "arXiv",
    doi = "10.1103/PhysRevLett.89.151802",
    journal = "Phys. Rev. Lett.",
    volume = "89",
    pages = "151802",
    year = "2002"
}

@article{BaBar:2005pon,
    author = "Aubert, Bernard and others",
    collaboration = "BaBar",
    title = "{A Study of $e^{+} e^{-} \to p \bar{p}$ using initial state radiation with BABAR}",
    eprint = "hep-ex/0512023",
    archivePrefix = "arXiv",
    reportNumber = "SLAC-PUB-11587, BABAR-PUB-05-050",
    doi = "10.1103/PhysRevD.73.012005",
    journal = "Phys. Rev. D",
    volume = "73",
    pages = "012005",
    year = "2006"
}

@article{BESIII:2011aa,
    author = "Ablikim, M. and others",
    collaboration = "BESIII",
    title = "{Spin-Parity Analysis of $p\bar{p}$ Mass Threshold Structure in $J/\psi$ and $\psi^\prime$ Radiative Decays}",
    eprint = "1112.0942",
    archivePrefix = "arXiv",
    primaryClass = "hep-ex",
    doi = "10.1103/PhysRevLett.108.112003",
    journal = "Phys. Rev. Lett.",
    volume = "108",
    pages = "112003",
    year = "2012"
}

@article{BaBar:2005sdl,
    author = "Aubert, Bernard and others",
    collaboration = "BaBar",
    title = "{Measurement of the $B^+ \to p \bar{p} K^+$ branching fraction and study of the decay dynamics}",
    eprint = "hep-ex/0507012",
    archivePrefix = "arXiv",
    reportNumber = "BABAR-PUB-05-20, SLAC-PUB-11296, BABAR-PUB-05-020",
    doi = "10.1103/PhysRevD.72.051101",
    journal = "Phys. Rev. D",
    volume = "72",
    pages = "051101",
    year = "2005"
}

@article{Kang:2013uia,
    author = "Kang, Xian-Wei and Haidenbauer, Johann and Mei\ss{}ner, Ulf-G.",
    title = "{Antinucleon-nucleon interaction in chiral effective field theory}",
    eprint = "1311.1658",
    archivePrefix = "arXiv",
    primaryClass = "hep-ph",
    doi = "10.1007/JHEP02(2014)113",
    journal = "JHEP",
    volume = "02",
    pages = "113",
    year = "2014"
}

@article{Dai:2017ont,
    author = "Dai, Ling-Yun and Haidenbauer, Johann and Mei\ss{}ner, Ulf-G",
    title = "{Antinucleon-nucleon interaction at next-to-next-to-next-to-leading order in chiral effective field theory}",
    eprint = "1702.02065",
    archivePrefix = "arXiv",
    primaryClass = "nucl-th",
    doi = "10.1007/JHEP07(2017)078",
    journal = "JHEP",
    volume = "07",
    pages = "078",
    year = "2017"
}

@article{Dover:1980pd,
    author = "Dover, C. B. and Richard, J. M.",
    title = "{Elastic, Charge Exchange, and Inelastic anti-p p Cross-Sections in the Optical Model}",
    reportNumber = "BNL-26766",
    doi = "10.1103/PhysRevC.21.1466",
    journal = "Phys. Rev. C",
    volume = "21",
    pages = "1466--1471",
    year = "1980"
}

@article{Dover:1981pp,
    author = "Dover, C. B. and Richard, J. M.",
    title = "{Spin Observables in Low-energy Nucleon Anti-nucleon Scattering}",
    reportNumber = "BNL-30271",
    doi = "10.1103/PhysRevC.25.1952",
    journal = "Phys. Rev. C",
    volume = "25",
    pages = "1952--1958",
    year = "1982"
}

@article{Cote:1982gr,
    author = "Cote, J. and Lacombe, M. and Loiseau, B. and Moussallam, B. and Vinh Mau, R.",
    title = "{On the Nucleon - anti-Nucleon Optical Potential}",
    reportNumber = "IPNO/TH 82-8",
    doi = "10.1103/PhysRevLett.48.1319",
    journal = "Phys. Rev. Lett.",
    volume = "48",
    pages = "1319",
    year = "1982"
}

@article{Hippchen:1991rr,
    author = "Hippchen, T. and Haidenbauer, J. and Holinde, K. and Mull, V.",
    title = "{Meson - baryon dynamics in the nucleon - anti-nucleon system. 1. The Nucleon - anti-nucleon interaction}",
    reportNumber = "KFA-IKP-TH-1991-3",
    doi = "10.1103/PhysRevC.44.1323",
    journal = "Phys. Rev. C",
    volume = "44",
    pages = "1323--1336",
    year = "1991"
}

@article{Mull:1994gz,
    author = "Mull, V. and Holinde, K.",
    title = "{Combined description of anti-N N scattering and annihilation with a hadronic model}",
    eprint = "nucl-th/9411014",
    archivePrefix = "arXiv",
    reportNumber = "KFA-IKP-TH-1994-39",
    doi = "10.1103/PhysRevC.51.2360",
    journal = "Phys. Rev. C",
    volume = "51",
    pages = "2360--2371",
    year = "1995"
}

@article{Zhou:2012ui,
    author = "Zhou, Daren and Timmermans, Rob G. E.",
    title = "{Energy-dependent partial-wave analysis of all antiproton-proton scattering data below 925 MeV/c}",
    eprint = "1210.7074",
    archivePrefix = "arXiv",
    primaryClass = "hep-ph",
    doi = "10.1103/PhysRevC.86.044003",
    journal = "Phys. Rev. C",
    volume = "86",
    pages = "044003",
    year = "2012"
}

@article{Dai:2018tlc,
    author = "Dai, Ling-Yun and Haidenbauer, Johann and Mei\ss{}ner, Ulf-G.",
    title = "{$J/\psi \to \gamma\eta'\pi^+\pi^-$ and the structure observed around the $\bar pp$ threshold}",
    eprint = "1804.07077",
    archivePrefix = "arXiv",
    primaryClass = "hep-ph",
    doi = "10.1103/PhysRevD.98.014005",
    journal = "Phys. Rev. D",
    volume = "98",
    number = "1",
    pages = "014005",
    year = "2018"
}

@article{Haidenbauer:2014kja,
    author = "Haidenbauer, J. and Kang, X. -W. and Mei\ss{}ner, U. -G.",
    title = "{The electromagnetic form factors of the proton in the timelike region}",
    eprint = "1405.1628",
    archivePrefix = "arXiv",
    primaryClass = "nucl-th",
    doi = "10.1016/j.nuclphysa.2014.06.007",
    journal = "Nucl. Phys. A",
    volume = "929",
    pages = "102--118",
    year = "2014"
}

@article{Yang:2022qoy,
    author = "Yang, Qin-He and Guo, Di and Dai, Ling-Yun and Haidenbauer, Johann and Kang, Xian-Wei and Mei\ss{}ner, Ulf-G.",
    title = "{New insights into the oscillations of the nucleon electromagnetic form factors}",
    eprint = "2206.01494",
    archivePrefix = "arXiv",
    primaryClass = "nucl-th",
    doi = "10.1016/j.scib.2023.09.036",
    journal = "Sci. Bull.",
    volume = "68",
    pages = "2729--2733",
    year = "2023"
}

@article{Zhou:2022loi,
    author = "Zhou, Daren and Long, Bingwei and Timmermans, R. G. E. and van Kolck, U.",
    title = "{Renormalization of one-pion exchange in chiral effective field theory for antinucleon-nucleon scattering}",
    eprint = "2203.06840",
    archivePrefix = "arXiv",
    primaryClass = "nucl-th",
    doi = "10.1103/PhysRevC.105.054005",
    journal = "Phys. Rev. C",
    volume = "105",
    number = "5",
    pages = "054005",
    year = "2022"
}

@article{Geng:2008mf,
    author = "Geng, L. S. and Martin Camalich, J. and Alvarez-Ruso, L. and Vicente Vacas, M. J.",
    title = "{Leading SU(3)-breaking corrections to the baryon magnetic moments in Chiral Perturbation Theory}",
    eprint = "0805.1419",
    archivePrefix = "arXiv",
    primaryClass = "hep-ph",
    doi = "10.1103/PhysRevLett.101.222002",
    journal = "Phys. Rev. Lett.",
    volume = "101",
    pages = "222002",
    year = "2008"
}

@article{Alarcon:2011zs,
    author = "Alarcon, J. M. and Martin Camalich, J. and Oller, J. A.",
    title = "{The chiral representation of the $\pi N$ scattering amplitude and the pion-nucleon sigma term}",
    eprint = "1110.3797",
    archivePrefix = "arXiv",
    primaryClass = "hep-ph",
    doi = "10.1103/PhysRevD.85.051503",
    journal = "Phys. Rev. D",
    volume = "85",
    pages = "051503",
    year = "2012"
}

@article{Ren:2012aj,
    author = "Ren, X. -L. and Geng, L. S. and Martin Camalich, J. and Meng, J. and Toki, H.",
    title = "{Octet baryon masses in next-to-next-to-next-to-leading order covariant baryon chiral perturbation theory}",
    eprint = "1209.3641",
    archivePrefix = "arXiv",
    primaryClass = "nucl-th",
    doi = "10.1007/JHEP12(2012)073",
    journal = "JHEP",
    volume = "12",
    pages = "073",
    year = "2012"
}

@article{Ren:2017yvw,
    author = "Ren, Xiu-Lei and Wang, Chun-Xuan and Li, Kai-Wen and Geng, Li-Sheng and Meng, Jie",
    title = "{Relativistic Chiral Description of the 1 S 0 Nucleon\textendash{}Nucleon Scattering}",
    eprint = "1712.10083",
    archivePrefix = "arXiv",
    primaryClass = "nucl-th",
    doi = "10.1088/0256-307X/38/6/062101",
    journal = "Chin. Phys. Lett.",
    volume = "38",
    number = "6",
    pages = "062101",
    year = "2021"
}

@article{Xiao:2020ozd,
    author = "Xiao, Yang and Wang, Chun-Xuan and Lu, Jun-Xu and Geng, Li-Sheng",
    title = "{Two-pion exchange contributions to the nucleon-nucleon interaction in covariant baryon chiral perturbation theory}",
    eprint = "2007.13675",
    archivePrefix = "arXiv",
    primaryClass = "nucl-th",
    doi = "10.1103/PhysRevC.102.054001",
    journal = "Phys. Rev. C",
    volume = "102",
    number = "5",
    pages = "054001",
    year = "2020"
}

@article{Wang:2021kos,
    author = "Wang, Chun-Xuan and Lu, Jun-Xu and Xiao, Yang and Geng, Li-Sheng",
    title = "{Nonperturbative two-pion exchange contributions to the nucleon-nucleon interaction in covariant baryon chiral perturbation theory}",
    eprint = "2110.05278",
    archivePrefix = "arXiv",
    primaryClass = "nucl-th",
    doi = "10.1103/PhysRevC.105.014003",
    journal = "Phys. Rev. C",
    volume = "105",
    number = "1",
    pages = "014003",
    year = "2022"
}

@article{Ren:2016jna,
    author = "Ren, Xiu-Lei and Li, Kai-Wen and Geng, Li-Sheng and Long, Bing-Wei and Ring, Peter and Meng, Jie",
    title = "{Leading order relativistic chiral nucleon-nucleon interaction}",
    eprint = "1611.08475",
    archivePrefix = "arXiv",
    primaryClass = "nucl-th",
    reportNumber = "CTP-SCU-2016012",
    doi = "10.1088/1674-1137/42/1/014103",
    journal = "Chin. Phys. C",
    volume = "42",
    number = "1",
    pages = "014103",
    year = "2018"
}

@article{Xiao:2018jot,
    author = "Xiao, Yang and Geng, Li-Sheng and Ren, Xiu-Lei",
    title = "{Covariant chiral nucleon-nucleon contact Lagrangian up to order $\mathcal{O}(q^4)$}",
    eprint = "1812.03005",
    archivePrefix = "arXiv",
    primaryClass = "nucl-th",
    doi = "10.1103/PhysRevC.99.024004",
    journal = "Phys. Rev. C",
    volume = "99",
    number = "2",
    pages = "024004",
    year = "2019"
}

@article{Lu:2021gsb,
    author = "Lu, Jun-Xu and Wang, Chun-Xuan and Xiao, Yang and Geng, Li-Sheng and Meng, Jie and Ring, Peter",
    title = "{Accurate Relativistic Chiral Nucleon-Nucleon Interaction up to Next-to-Next-to-Leading Order}",
    eprint = "2111.07766",
    archivePrefix = "arXiv",
    primaryClass = "nucl-th",
    doi = "10.1103/PhysRevLett.128.142002",
    journal = "Phys. Rev. Lett.",
    volume = "128",
    number = "14",
    pages = "142002",
    year = "2022"
}

@article{Lu:2022hwm,
    author = "Lu, Jun-Xu and Geng, Li-Sheng and Doering, Michael and Mai, Maxim",
    title = "{Cross-Channel Constraints on Resonant Antikaon-Nucleon Scattering}",
    eprint = "2209.02471",
    archivePrefix = "arXiv",
    primaryClass = "hep-ph",
    doi = "10.1103/PhysRevLett.130.071902",
    journal = "Phys. Rev. Lett.",
    volume = "130",
    number = "7",
    pages = "071902",
    year = "2023"
}

@article{BES:2005ega,
    author = "Ablikim, M. and others",
    collaboration = "BES",
    title = "{Observation of a resonance X(1835) in J / psi -\ensuremath{>} gamma pi+ pi- eta-prime}",
    eprint = "hep-ex/0508025",
    archivePrefix = "arXiv",
    doi = "10.1103/PhysRevLett.95.262001",
    journal = "Phys. Rev. Lett.",
    volume = "95",
    pages = "262001",
    year = "2005"
}

@article{BESIII:2016fbr,
    author = "Ablikim, M. and others",
    collaboration = "BESIII",
    title = "{Observation of an anomalous line shape of the $\eta^{\prime}\pi^{+}\pi^{-}$ mass spectrum near the $p\bar{p}$ mass threshold in $J/\psi\rightarrow\gamma\eta^{\prime}\pi^{+}\pi^{-}$}",
    eprint = "1603.09653",
    archivePrefix = "arXiv",
    primaryClass = "hep-ex",
    doi = "10.1103/PhysRevLett.117.042002",
    journal = "Phys. Rev. Lett.",
    volume = "117",
    number = "4",
    pages = "042002",
    year = "2016"
}

@article{BESIII:2021xoh,
    author = "Ablikim, M. and others",
    collaboration = "BESIII",
    title = "{Observation of $J/\psi$ Electromagnetic Dalitz Decays to $X(1835)$, $X(2120)$ and $X(2370)$}",
    eprint = "2112.14369",
    archivePrefix = "arXiv",
    primaryClass = "hep-ex",
    doi = "10.1103/PhysRevLett.129.022002",
    journal = "Phys. Rev. Lett.",
    volume = "129",
    number = "2",
    pages = "022002",
    year = "2022"
}

@article{Machleidt:2011zz,
    author = "Machleidt, R. and Entem, D. R.",
    title = "{Chiral effective field theory and nuclear forces}",
    eprint = "1105.2919",
    archivePrefix = "arXiv",
    primaryClass = "nucl-th",
    doi = "10.1016/j.physrep.2011.02.001",
    journal = "Phys. Rept.",
    volume = "503",
    pages = "1--75",
    year = "2011"
}

@article{Baru:2011bw,
    author = "Baru, V. and Hanhart, C. and Hoferichter, M. and Kubis, B. and Nogga, A. and Phillips, D. R.",
    title = "{Precision calculation of threshold $\pi^- d$ scattering, $\pi N$ scattering lengths, and the GMO sum rule}",
    eprint = "1107.5509",
    archivePrefix = "arXiv",
    primaryClass = "nucl-th",
    doi = "10.1016/j.nuclphysa.2011.09.015",
    journal = "Nucl. Phys. A",
    volume = "872",
    pages = "69--116",
    year = "2011"
}

@article{Chen:2012nx,
    author = "Chen, Yun-Hua and Yao, De-Liang and Zheng, H. Q.",
    title = "{Analyses of pion-nucleon elastic scattering amplitudes up to $O(p^4)$ in the extended-on-mass-shell subtraction scheme}",
    eprint = "1212.1893",
    archivePrefix = "arXiv",
    primaryClass = "hep-ph",
    doi = "10.1103/PhysRevD.87.054019",
    journal = "Phys. Rev. D",
    volume = "87",
    number = "5",
    pages = "054019",
    year = "2013"
}

@article{Nieves:2003in,
    author = "Nieves, Jose F. and Pal, Palash B.",
    title = "{Generalized Fierz identities}",
    eprint = "hep-ph/0306087",
    archivePrefix = "arXiv",
    reportNumber = "SINP-TNP-03-16",
    doi = "10.1119/1.1757445",
    journal = "Am. J. Phys.",
    volume = "72",
    pages = "1100--1108",
    year = "2004"
}

@article{Stapp:1956mz,
    author = "Stapp, H. P. and Ypsilantis, T. J. and Metropolis, N.",
    title = "{Phase shift analysis of 310-MeV proton proton scattering experiments}",
    doi = "10.1103/PhysRev.105.302",
    journal = "Phys. Rev.",
    volume = "105",
    pages = "302--310",
    year = "1957"
}

@article{BESIII:2023vvr,
    author = "Ablikim, M. and others",
    collaboration = "BESIII",
    title = "{Observation of the Anomalous Shape of $X(1840)$ in $J/\psi\to\gamma 3(\pi^+\pi^-)$ Indicating a Second Resonance Near the $p\bar p$ Threshold}",
    eprint = "2310.17937",
    archivePrefix = "arXiv",
    primaryClass = "hep-ex",
    doi = "10.1103/PhysRevLett.132.151901",
    journal = "Phys. Rev. Lett.",
    volume = "132",
    number = "15",
    pages = "151901",
    year = "2024"
}

@article{Zou:2023quo,
    author = "Zou, Wei-Jiang and Lu, Jun-Xu and Zhao, Peng-Wei and Geng, Li-Sheng and Meng, Jie",
    title = "{Saturation of nuclear matter in the relativistic Brueckner-Hatree-Fock approach with a leading order covariant chiral nuclear force}",
    eprint = "2312.15672",
    archivePrefix = "arXiv",
    primaryClass = "nucl-th",
    doi = "10.1016/j.physletb.2024.138732",
    journal = "Phys. Lett. B",
    volume = "854",
    pages = "138732",
    year = "2024"
}

@article{Dedonder:2009bk,
    author = "Dedonder, J. -P. and Loiseau, B. and El-Bennich, B. and Wycech, S.",
    title = "{On the structure of the X(1835) baryonium}",
    eprint = "0904.2163",
    archivePrefix = "arXiv",
    primaryClass = "nucl-th",
    reportNumber = "LPNHE-2009-03",
    doi = "10.1103/PhysRevC.80.045207",
    journal = "Phys. Rev. C",
    volume = "80",
    pages = "045207",
    year = "2009"
}

@article{Sibirtsev:2004id,
    author = "Sibirtsev, Alexander and Haidenbauer, Johann and Krewald, S. and Meissner, Ulf-G. and Thomas, Anthony William",
    title = "{Near threshold enhancement of the p anti-p mass spectrum in j/psi decay}",
    eprint = "hep-ph/0411386",
    archivePrefix = "arXiv",
    reportNumber = "FZJ-IKP-TH-2004-20, HISKP-TH-04-24, JLAB-THY-05-01",
    doi = "10.1103/PhysRevD.71.054010",
    journal = "Phys. Rev. D",
    volume = "71",
    pages = "054010",
    year = "2005"
}

@article{Yan:2004xs,
    author = "Yan, Mu-Lin and Li, Si and Wu, Bin and Ma, Bo-Qiang",
    title = "{Nucleon-antinucleon bound states in Skyrmion-type potential}",
    eprint = "hep-ph/0405087",
    archivePrefix = "arXiv",
    doi = "10.1103/PhysRevD.72.034027",
    journal = "Phys. Rev. D",
    volume = "72",
    pages = "034027",
    year = "2005"
}

@article{Xiao:2024nbarN,
    author = "Xiao, Yang and Lu, Jun-Xu and Geng, Li-Sheng",
    title = "{Reexamination of antinucleon-nucleon interactions in covariant chiral effective field theory}",
    eprint = "2406.01292",
    archivePrefix = "arXiv",
    primaryClass = "nucl-th",
    doi = "10.1103/PhysRevC.110.064002",
    journal = "Phys. Rev. C",
    volume = "110",
    number = "6",
    pages = "064002",
    year = "2024"
}

@article{LHCb:2025ppbar,
    author = "Aaij, R. and others",
    collaboration = "LHCb",
    title = "{First observation of the charmless baryonic decay $B^+\to\bar{\Lambda}p\bar{p}p$}",
    eprint = "2508.16259",
    archivePrefix = "arXiv",
    primaryClass = "hep-ex",
    reportNumber = "LHCb-PAPER-2025-032, CERN-EP-2025-174",
    doi = "10.1103/3pcs-dxtn",
    journal = "Phys. Rev. Lett.",
    volume = "135",
    number = "26",
    pages = "261901",
    year = "2025"
}

@article{Niu:2024pole,
    author = "Niu, Peng-Yu and Zhang, Zhen-Yu and Li, Yi-Yao and Wang, Qian and Zhao, Qiang",
    title = "{The pole structures of the $X(1840)/X(1835)$ and the $X(1880)$}",
    eprint = "2408.14876",
    archivePrefix = "arXiv",
    primaryClass = "hep-ph",
    doi = "10.1103/PhysRevD.110.094020",
    journal = "Phys. Rev. D",
    volume = "110",
    number = "9",
    pages = "094020",
    year = "2024"
}

@article{Yang:2024x1880,
    author = "Yang, Qin-He and Dai, Ling-Yun and Mei{\ss}ner, Ulf-G.",
    title = "{Quantum numbers of the $X(1880)$}",
    eprint = "2412.07599",
    archivePrefix = "arXiv",
    primaryClass = "hep-ph",
    doi = "10.1103/gmyw-cfjf",
    journal = "Phys. Rev. D",
    volume = "113",
    number = "1",
    pages = "L011501",
    year = "2026"
}

@article{Zheng:2025lambdaN,
    author = "Zheng, Ru-You and Liu, Zhi-Wei and Geng, Li-Sheng and Hu, Jin-Niu and Wang, Sibo",
    title = "{In-medium $\Lambda N$ interactions with leading-order covariant chiral hyperon/nucleon-nucleon forces}",
    eprint = "2501.02826",
    archivePrefix = "arXiv",
    primaryClass = "nucl-th",
    doi = "10.1016/j.physletb.2025.139416",
    journal = "Phys. Lett. B",
    volume = "864",
    pages = "139416",
    year = "2025"
}

@article{Kadyshevsky:1968,
    author = "Kadyshevsky, V. G.",
    title = "{Quasipotential type equation for the relativistic scattering amplitude}",
    doi = "10.1016/0550-3213(68)90274-5",
    journal = "Nucl. Phys. B",
    volume = "6",
    number = "2",
    pages = "125--148",
    year = "1968"
}

@article{Furnstahl:2015rha,
    author = "Furnstahl, R. J. and Klco, N. and Phillips, D. R. and Wesolowski, S.",
    title = "{Quantifying truncation errors in effective field theory}",
    eprint = "1506.01343",
    archivePrefix = "arXiv",
    primaryClass = "nucl-th",
    doi = "10.1103/PhysRevC.92.024005",
    journal = "Phys. Rev. C",
    volume = "92",
    number = "2",
    pages = "024005",
    year = "2015"
}

@article{Melendez:2017phj,
    author = "Melendez, J. A. and Wesolowski, S. and Furnstahl, R. J.",
    title = "{Bayesian truncation errors in chiral effective field theory: Nucleon-nucleon observables}",
    eprint = "1704.03308",
    archivePrefix = "arXiv",
    primaryClass = "nucl-th",
    doi = "10.1103/PhysRevC.96.024003",
    journal = "Phys. Rev. C",
    volume = "96",
    number = "2",
    pages = "024003",
    year = "2017"
}

@article{Melendez:2019izc,
    author = "Melendez, J. A. and Furnstahl, R. J. and Phillips, D. R. and Pratola, M. T. and Wesolowski, S.",
    title = "{Quantifying correlated truncation errors in effective field theory}",
    eprint = "1904.10581",
    archivePrefix = "arXiv",
    primaryClass = "nucl-th",
    doi = "10.1103/PhysRevC.100.044001",
    journal = "Phys. Rev. C",
    volume = "100",
    number = "4",
    pages = "044001",
    year = "2019"
}

@article{Xiao:2025ufy,
    author = "Xiao, Yang and Lu, Jun-Xu and Song, Chun-Yan and Geng, Li-Sheng",
    title = "{Charge-dependent nucleon-nucleon potentials in covariant chiral effective field theory}",
    eprint = "2504.15598",
    archivePrefix = "arXiv",
    primaryClass = "nucl-th",
    doi = "10.1103/85cd-6gzn",
    journal = "Phys. Rev. C",
    volume = "114",
    number = "1",
    pages = "014002",
    year = "2026"
}

@article{Zou:2025dao,
    author = "Zou, Wei-Jiang and Yang, Yi-Long and Lu, Jun-Xu and Zhao, Peng-Wei and Geng, Li-Sheng and Meng, Jie",
    title = "{Nuclear and neutron matter in the relativistic Brueckner-Hartree-Fock theory with a next-to-leading-order covariant chiral nuclear force}",
    eprint = "2506.18519",
    archivePrefix = "arXiv",
    primaryClass = "nucl-th",
    doi = "10.1103/21c6-xvh6",
    journal = "Phys. Rev. C",
    volume = "113",
    number = "1",
    pages = "014311",
    year = "2026"
}

@article{Kang:2015yka,
    author = "Kang, Xian-Wei and Haidenbauer, J. and Mei{\ss}ner, Ulf-G.",
    title = "{Near-threshold $\bar pp$ invariant mass spectrum measured in $J/\psi$ and $\psi'$ decays}",
    eprint = "1502.00880",
    archivePrefix = "arXiv",
    primaryClass = "nucl-th",
    doi = "10.1103/PhysRevD.91.074003",
    journal = "Phys. Rev. D",
    volume = "91",
    number = "7",
    pages = "074003",
    year = "2015"
}

@article{Haidenbauer:2015yka,
    author = "Haidenbauer, Johann and Hanhart, Christoph and Kang, Xian-Wei and Mei{\ss}ner, Ulf-G.",
    title = "{Origin of the structures observed in $e^+e^-$ annihilation into multipion states around the $\bar pp$ threshold}",
    eprint = "1506.08120",
    archivePrefix = "arXiv",
    primaryClass = "nucl-th",
    doi = "10.1103/PhysRevD.92.054032",
    journal = "Phys. Rev. D",
    volume = "92",
    number = "5",
    pages = "054032",
    year = "2015"
}

@article{Yang:2024iuc,
    author = "Yang, Qin-He and Guo, Di and Li, Ming-Yan and Dai, Ling-Yun and Haidenbauer, Johann and Mei{\ss}ner, Ulf-G.",
    title = "{Study of the electromagnetic form factors of the nucleons in the timelike region}",
    eprint = "2404.12448",
    archivePrefix = "arXiv",
    primaryClass = "nucl-th",
    doi = "10.1007/JHEP08(2024)208",
    journal = "JHEP",
    volume = "08",
    pages = "208",
    year = "2024"
}

@article{Ji:2026mfe,
    author = "Ji, Teng and Mei{\ss}ner, Ulf-G.",
    title = "{Understanding the near-threshold structures in $e^+e^-$ annihilation from a unified $N\bar{N}$-interaction perspective}",
    eprint = "2607.06168",
    archivePrefix = "arXiv",
    primaryClass = "hep-ph",
    doi = "10.1140/epjc/s10052-026-16364-x",
    journal = "Eur. Phys. J. C",
    volume = "86",
    number = "9",
    pages = "1116",
    year = "2026"
}

@article{Chen:2010NbarNEFT,
  author        = {Chen, G. Y. and Dong, H. R. and Ma, J. P.},
  title         = {Near threshold enhancement of {$p\bar{p}$} system and {$p\bar{p}$} elastic scattering},
  journal       = {Phys. Lett. B},
  volume        = {692},
  pages         = {136--142},
  year          = {2010},
  doi           = {10.1016/j.physletb.2010.07.030},
  eprint        = {1004.5174},
  archivePrefix = {arXiv},
  primaryClass  = {hep-ph}
}

@article{ChenMa:2011NLO,
  author        = {Chen, G. Y. and Ma, J. P.},
  title         = {{$N\bar{N}$} scattering at next-to-leading order in an effective theory},
  journal       = {Phys. Rev. D},
  volume        = {83},
  pages         = {094029},
  year          = {2011},
  doi           = {10.1103/PhysRevD.83.094029},
  eprint        = {1101.4071},
  archivePrefix = {arXiv},
  primaryClass  = {hep-ph}
}

@article{ElBennich:2009Paris,
  author        = {El-Bennich, B. and Lacombe, M. and Loiseau, B. and Wycech, S.},
  title         = {Paris {$N\bar{N}$} potential constrained by recent antiprotonic-atom data and {$\bar{n}p$} total cross sections},
  journal       = {Phys. Rev. C},
  volume        = {79},
  pages         = {054001},
  year          = {2009},
  doi           = {10.1103/PhysRevC.79.054001},
  eprint        = {0807.4454},
  archivePrefix = {arXiv},
  primaryClass  = {nucl-th}
}

@article{Fuchs:2003EOMS,
  author        = {Fuchs, T. and Gegelia, J. and Japaridze, G. and Scherer, S.},
  title         = {Renormalization of relativistic baryon chiral perturbation theory and power counting},
  journal       = {Phys. Rev. D},
  volume        = {68},
  pages         = {056005},
  year          = {2003},
  doi           = {10.1103/PhysRevD.68.056005},
  eprint        = {hep-ph/0302117},
  archivePrefix = {arXiv}
}

@article{Zhou:2023HigherWaves,
  author        = {Zhou, Daren},
  title         = {Renormalization of one-pion exchange in higher partial waves in chiral effective field theory for antinucleon-nucleon system},
  journal       = {Chin. Phys. C},
  volume        = {47},
  number        = {3},
  pages         = {034103},
  year          = {2023},
  doi           = {10.1088/1674-1137/aca467},
  eprint        = {2210.11683},
  archivePrefix = {arXiv},
  primaryClass  = {nucl-th}
}

@article{ZouChiang:2004OPEFSI,
  author        = {Zou, B. S. and Chiang, H. C.},
  title         = {One-pion-exchange final-state interaction and the {$p\bar{p}$} near threshold enhancement in {$J/\psi\to\gamma p\bar{p}$} decays},
  journal       = {Phys. Rev. D},
  volume        = {69},
  pages         = {034004},
  year          = {2004},
  doi           = {10.1103/PhysRevD.69.034004},
  eprint        = {hep-ph/0309273},
  archivePrefix = {arXiv}
}

@article{Kerbikov:2004Threshold,
  author        = {Kerbikov, B. and Stavinsky, A. and Fedotov, V.},
  title         = {Model-independent view on the low-mass proton-antiproton enhancement},
  journal       = {Phys. Rev. C},
  volume        = {69},
  pages         = {055205},
  year          = {2004},
  doi           = {10.1103/PhysRevC.69.055205},
  eprint        = {hep-ph/0402054},
  archivePrefix = {arXiv}
}

@article{DattaODonnell:2003Baryonium,
  author        = {Datta, Alakabha and O'Donnell, Patrick J.},
  title         = {A new state of baryonium},
  journal       = {Phys. Lett. B},
  volume        = {567},
  pages         = {273--276},
  year          = {2003},
  doi           = {10.1016/j.physletb.2003.06.050},
  eprint        = {hep-ph/0306097},
  archivePrefix = {arXiv}
}

@article{Haidenbauer:2006Timelike,
  author        = {Haidenbauer, J. and Hammer, H.-W. and Mei{\ss}ner, Ulf-G. and Sibirtsev, A.},
  title         = {On the strong energy dependence of the {$e^+e^-\leftrightarrow p\bar{p}$} amplitude near threshold},
  journal       = {Phys. Lett. B},
  volume        = {643},
  pages         = {29--32},
  year          = {2006},
  doi           = {10.1016/j.physletb.2006.10.016},
  eprint        = {hep-ph/0606064},
  archivePrefix = {arXiv}
}

@article{EntemFernandez:2006QuarkModel,
  author        = {Entem, D. R. and Fern{\'a}ndez, F.},
  title         = {The {$N\bar{N}$} interaction in a constituent quark model: Baryonium states and protonium level shifts},
  journal       = {Phys. Rev. C},
  volume        = {73},
  pages         = {045214},
  year          = {2006},
  doi           = {10.1103/PhysRevC.73.045214}
}

@article{Dmitriev:2016RealVirtual,
  author        = {Dmitriev, V. F. and Milstein, A. I. and Salnikov, S. G.},
  title         = {Real and virtual {$N\bar{N}$} pair production near the threshold},
  journal       = {Phys. Rev. D},
  volume        = {93},
  pages         = {034033},
  year          = {2016},
  doi           = {10.1103/PhysRevD.93.034033},
  eprint        = {1512.00951},
  archivePrefix = {arXiv},
  primaryClass  = {hep-ph}
}

@article{MilsteinSalnikov:2017Jpsi,
  author        = {Milstein, A. I. and Salnikov, S. G.},
  title         = {Interaction of real and virtual {$p\bar{p}$} pairs in {$J/\psi\to p\bar{p}\gamma(\rho,\omega)$} decays},
  journal       = {Nucl. Phys. A},
  volume        = {966},
  pages         = {54--63},
  year          = {2017},
  doi           = {10.1016/j.nuclphysa.2017.06.002},
  eprint        = {1701.05741},
  archivePrefix = {arXiv},
  primaryClass  = {hep-ph}
}

@article{Zeng:2025fqb,
    author = "Zeng, Xuan and Zheng, Ru-You and Liu, Zhi-Wei and Geng, Li-Sheng and Zhou, Xian-Rong",
    title = "{Hypernuclear structure with a new leading-order covariant chiral hyperon-nucleon force}",
    eprint = "2509.16878",
    archivePrefix = "arXiv",
    primaryClass = "nucl-th",
    doi = "10.1103/9s7s-ypl3",
    journal = "Phys. Rev. C",
    volume = "113",
    number = "1",
    pages = "014310",
    year = "2026"
}

@article{Shen:2025iue,
    author = "Shen, Shihang and Lu, Jun-Xu and Geng, Li-Sheng and Zou, Wei-Jiang and Meng, Jie",
    title = "{From bare two-nucleon interaction to nuclear matter and finite nuclei in a relativistic framework}",
    eprint = "2507.01257",
    archivePrefix = "arXiv",
    primaryClass = "nucl-th",
    doi = "10.1103/pd1h-cdwc",
    journal = "Phys. Rev. C",
    volume = "114",
    number = "2",
    pages = "024328",
    year = "2026"
}

@article{Lu:2025review,
  author  = {Lu, Jun-Xu and Xiao, Yang and Liu, Zhi-Wei and Geng, Li-Sheng},
  title   = {Relativistic chiral nuclear forces: Status and prospects},
  journal = {Int. J. Mod. Phys. E},
  volume  = {34},
  pages   = {2543007},
  year    = {2025}
}

@article{Lu:2024review,
  author  = {Lu, Jun-Xu and Xiao, Yang and Liu, Zhi-Wei and Geng, Li-Sheng},
  title   = {New Generation of High Precision Nuclear Forces---Relativistic Chiral Nuclear Force---Status and Prospects},
  journal = {Nucl. Phys. Rev.},
  volume  = {41},
  pages   = {35--50},
  year    = {2024}
}

@misc{Yang:2026barB,
    author = "Yang, Qin-He and Haidenbauer, Johann and Mei{\ss}ner, Ulf-G.",
    title = "{The antibaryon-baryon interaction based on chiral EFT with a focus on $\bar pp$ and $\bar\Lambda\Lambda$ reactions}",
    eprint = "2611.26357",
    archivePrefix = "arXiv",
    primaryClass = "nucl-th",
    year = "2026"
}
\end{document}